\documentclass{aa}  

\usepackage{amsfonts}
\usepackage{amsmath}
\usepackage{bm}
\usepackage{xcolor}
\usepackage{mathrsfs}
\usepackage{graphicx}
\usepackage[varg]{txfonts}
\usepackage[colorlinks=true, allcolors=blue]{hyperref}
\begin{document}

   \title{Exact hybrid-kinetic equilibria for magnetized plasmas with shearing flows}


   \author{G. Guzzi \inst{1}
          \and
          A. Settino\inst{1,2}
          \and
          F. Valentini\inst{1}
          \and 
          F. Malara \inst{1}
          }

   \institute{Dipartimento di Fisica, Universit\`a della Calabria,
              ponte P. Bucci, cubo 31C, 87036, Rende (CS), Italy \label{inst1}\\
              \and
             IRF Swedish Institute of Space Physics, \AA ngstr\"omlaboratoriet, L\"agerhyddvs 1, Uppsala, Sweden \label{inst2}\\
             }

   \date{Received xx; accepted xx}

  \abstract
   {Magnetized plasmas characterized by shearing flows are present in many natural contexts, such as the Earth's magnetopause and the solar wind. The collisionless nature of involved plasmas requires a kinetic description. When the width of the shear layer is of the order of ion scales, the Hybrid Vlasov-Maxwell approach can be adopted.}
   {The aim of the paper is to derive explicit forms for stationary configurations of magnetized plasmas with planar shearing flows, within the Hybrid Vlasov-Maxwell description. Two configurations are considered: the first with a uniform magnetic field obliquely directed with respect to the bulk velocity; and the second with a uniform-magnitude variable-direction magnetic field.}
   {Stationary ion distribution functions are obtained by combining single-particle constant of motions, which are derived studying particle dynamics. Preliminary information about the form of the distribution functions are analytically derived considering a local approximation for the background electromagnetic field. Then, a numerical method is set up to obtain a solution for general profiles.}
   {The explicit distribution functions that are found allow to obtain profiles of density, bulk velocity, temperature and heat flux. Anisotropy and agyrotropy in the distribution function are also evaluated. Stationarity of the solution during numerical simulations is checked in the uniform oblique magnetic field case.}
   {The considered configurations can be used as models for the Earth's magnetopause in simulations of the Kelvin-Helmholtz instability.}


   \maketitle
%

\section{Introduction}
Shearing flows in a fluid or in a plasma are characterized by a velocity field ${\bf u}$ that varies in a direction nearly perpendicular to ${\bf u}$. This kind of configuration is found in several astrophysical contexts; for instance, in the interaction region between fast and slow streams of the solar wind \citep{bruno13}; in the interface between the solar wind and a planetary \citep{sckopke81,fujimoto98,hasegawa04} or cometary magnetosphere \citep{ershkovich83,mccomas87,malara89}; in the solar corona at the surface limiting coronal mass ejections \citep{foullon11}; and in the interface between astrophysical jets and the surrounding medium \citep{hamlin13}. 
In a magnetized plasma a configuration with a shearing flow is unstable if the velocity jump $\Delta u$ across the shear layer is larger than a threshold, which is of the order of the Alfv\'en velocity associated with the magnetic field component $B_{||}$ parallel to ${\bf u}$. In this case a Kelvin-Helmholtz instability (KHI) develops (e.g., \citet{chandrasekhar61}).
KHI has been extensively studied at the Earth's magnetopause, where the solar wind plasma slips over a more static magnetosphere \citep{kivelson95,seon95,fairfield00,fairfield03,hasegawa04,hasegawa06,nykyri06}. There, the parallel magnetic field is small enough to allow for the development of the instability. In the Earth's  magnetosphere KHI determines several phenomena; for instance, it can favour the solar-wind plasma entry into the magnetosphere through magnetic reconnection that takes place in certain magnetic configurations \citep{eriksson16,nakamura17,sisti19}. 

From a theoretical point of view, the development of KHI in a plasma has been largely studied, either in the framework of Magnetohydrodynamics (MHD) \citep[e.g.,][]{axford60,walker81,miura82,contin03,matsumoto04,nakamura04,faganello08}, or using a kinetic description \citep{pritchett84,matsumoto06,cowee09,matsumoto10,nakamura10,nakamura11,nakamura13,henri13,karimabadi13}; for a review see also \citet{faganello17}. A kinetic approach is more appropriate than MHD in the case of the Earth's magnetopause, where the scale length associated with the shear layer is of the order of ion scales.
Stable shearing flows are relevant within the problem of wave propagation in nonuniform plasmas. When a perturbation propagates in an inhomogeneous velocity or magnetic field, fluctuations at small scales are generated leading to a fast wave dissipation. Moreover, couplings among different wave modes can take place. These effects have been widely studied both analytically and numerically using a MHD description \citep[e.g,][]{mok85,lee86,hollweg87,califano90,califano92,malara92,malara96a,kaghashvili99,tsiklauri02,landi05,kaghashvili06,kaghashvili07,pucci14}; while, in the framework of hybrid simulations, the study of propagation and interaction of Alfv\'enic wave packets in an inhomogeneous medium, have shown the importance of both collisions and kinetic effects, which lead to the generation of anisotropies and agyrotropies in the distribution function \citep[e.g.,][and references there in]{pezzi17a,pezzi17b}.
Such mechanisms of small-scale production have been invoked in wave-based models of coronal heating \citep[e.g.,][]{petkaki98,malara00,malara03,malara05,malara07}, as well as to describe turbulence evolution at the interface between fast and slow solar wind \citep{roberts92} or around the heliospheric current sheet \citep{malara96b}. More recently, the propagation of Alfv\'en waves on inhomogeneities at scales comparable with ion scales has been studied using both Hall-MHD and kinetic simulations \citep{vasconez15,pucci16,valentini17}, including a case when the wave propagates in a shearing flow \citep{maiorano20}. It has been shown that the wave evolution leads to formation of Kinetic Alfv\'en Waves, that are believed to play a role in energy dissipation for collisionless plasmas, as well as generation of field-aligned ion beams \citep{valentini17,maiorano20} similar to those observed in the solar wind \citep{marsch06}.

Prior to studying either KHI or the wave propagation in a shear layer, there is the problem of building up a stationary configuration representing the shearing flow. This is straighforward to do within a fluid theory. On the other hand, if the plasma is noncollisional and the shear width is comparable with ion scales -- as, for instance, at the magnetopause -- a kinetic approach is more suitable than a fluid one. In that case, building a stationary velocity-sheared configuration represents a much more complex task. In the fully kinetic case, particle distribution functions (DFs) corresponding to shearing flows has been found for a magnetic field parallel to the bulk velocity \citep{roytershteyn08}, a uniform perpendicular magnetic field \citep{ganguli88,nishikawa88,cai90}, and a nonuniform magnetic field \citep{mahajan00}. In spite of the existence of such exact stationary solutions, in many kinetic studies of the KHI a shifted Maxwellian (SM) has been used as unperturbed particle DF. However, since a SM is not an exact stationary state, this generates spurious oscillations that affect the development of the instability. In particular, \citet{settino20} have recently shown that both linear and nonlinear stages of KHI are different when an exact stationary solution is employed instead of a SM to represent the {\em same} shearing flow. Using an exact stationary background is essentially also in the problem of wave propagation in shearing flows \citep{maiorano20}, otherwise spurious oscillations would superpose over waves making difficult to follow the wave evolution.

In the present paper we deal with the problem of building up a stationary state representing a collisionless magnetized plasma with a shearing flow. We are interested in a situation where the length scale of the shear layer is of the order of ion scales (ion inertial length or Larmor radius). For this reason, we situate our study in the framework of the Hybrid Vlasov-Maxwell (HVM) theory. In the HVM model, ions are kinetically represented while electrons are treated as a massless fluid \citep{valentini07}. This kind of representation have been largely employed to describe phenomena at scales larger than, or of the order of ion scales \citep[e.g.,][]{cerri16,cerri17,franci15,matthaeus14,pezzi18,servidio12,servidio15, valentini11,valentini16,valentini17}. In particular, \citet{cerri13} have derived approximately stationary ion distribution functions, based on the evaluation of finite Larmor radius effects in the ion pressure tensor; their method has been employed to describe temperature anisotropy in the presence of shearing flows \citep{cerri14,delsarto16}. In a previous paper \citep[][hereafter Paper I]{malara18}, within the HVM framework, we have built up exact stationary configurations characterized by a shearing flow with a uniform magnetic field, either parallel or perpendicular to the plasma bulk velocity. The method is based on the determination of a stationary ion DF in terms of single-particle constants of motions. Such solutions have been employed to study the evolution of an Alfv\'enic disturbance in a shearing flow in the case of parallel magnetic field \citep{maiorano20}, as well as the development of KHI \citep{settino20} in the case of perpendicular magnetic field.

Here, we extend the stationary solutions found in Paper I to more general magnetized shearing flow configurations. Namely, a shearing flow with a uniform obliquely-directed magnetic field or variable-directed uniform-intensity magnetic field. Within the HVM approach, the present results can be employed to obtain a more realistic description of the KHI in the magnetopause. Indeed, the magnetic field across the magnetopause is neither exactly perpendicular to the plasma velocity nor exactly uniform, as assumed, for instance, in \citet{settino20}. Beside an improved realism, the present solutions have the further advantage to be exactly stationary as compared, for instance, with the approximate solutions by \citet{cerri13}. This latter aspect can affect the develpment of the instability, as shown by \citet{settino20}.

The plan of the paper is the following. In Section \ref{section2} the physical model is presented; in Section \ref{section3} the single-particle dynamics is described; in Section \ref{section4} a case with  a uniform magnetic field, obliquely directed with respect to the bulk velocity, is described; in Section \ref{section5} a case with uniform-intensity variable-direction magnetic field is discussed; finally, in Section \ref{section6} we draw the conclusions.

\section{Equations of the model}\label{section2}

We consider a collisionless plasma formed by protons and electrons. The HVM model \citep[e.g.,][]{valentini07} describes such a plasma at scales larger or of the order of proton scales (proton inertial length and/or Larmor radius). In this model protons are kinetically described by the particle DF $f({\bf x},{\bf v},t)$, where ${\bf x}$ and ${\bf v}$ are the position and the velocity coordinates in the phase-space, respectively, and $t$ is time. Electrons are treated as a massless fluid. Relevant moments of the proton DF are the density $n({\bf x},t) = \int f({\bf x},{\bf v},t)d^3{\bf v} $ and bulk velocity ${\bf u}({\bf x},t) = \int {\bf v} f({\bf x},{\bf v},t) d^3 {\bf v}/n({\bf x},t)$. Charge neutrality is assumed; therefore, the electron density is $n_e=n$. The proton DF evolves in time according to the Vlasov equation
\begin{equation}\label{vlasov}
\frac{\partial f}{\partial t} + {\bf v} \cdot \nabla f + \frac{e}{m} \left( {\bf E} + \frac{{\bf v}\times {\bf B}}{c} \right) \cdot \frac{\partial f}{\partial {\bf v}} = 0 
\end{equation}
where $e$ and $m$ are the proton charge and mass, ${\bf E}({\bf x},t)$ and ${\bf B}({\bf x},t)$ are the electric and magnetic fields, respectively, and $c$ is the speed of light. The set of equations includes the Faraday's law, the Ampere's law (where the displacement current is neglected) and the generalized Ohm's law:
\begin{equation}\label{far-amp}
\frac{\partial {\bf B}}{\partial t} = -c \nabla \times {\bf E} \;\; ; \;\; {\bf j} = \frac{c}{4\pi} \nabla \times {\bf B}
\end{equation}
\begin{equation}\label{ohm}
{\bf E}=-\frac{1}{c} {\bf u}\times {\bf B} + \frac{1}{en} \left( \frac{{\bf j}\times {\bf B}}{c} \right) -\frac{1}{en} \nabla p_e
\end{equation}
where ${\bf j}$ is the current density and $p_e$ is the electron pressure. In the HVM approach it is often assumed that the electron fluid is isothermal, so that the electron pressure $p_e$ is a function only of the density $n$. As shown later, in the present case we need to discard the isothermal assumption and treat $p_e$ as an independent quantity. Therefore a further equation is necessary to determine $p_e$. Here we assume that the electron fluid is adiabatic and include the equation:
\begin{equation}\label{adiab}
\left[ \frac{\partial}{\partial t} + \left( {\bf u}_e \cdot \nabla \right)\right] \left( \frac{p_e}{n^{\gamma_e}}\right) = 0
\end{equation}
where ${\bf u}_e = {\bf u} - {\bf j}/(en)$ and $\gamma_e$ are the electron bulk velocity and adiabatic index, respectively.

We look for a stationary solution of the system (\ref{vlasov})-(\ref{adiab}). If the electric and magnetic fields are constant in time, a DF expressed only in terms of single-particle constants of motion is a time-independent solution of the Vlasov equation (\ref{vlasov}). Therefore, to obtain a stationary solution we follow the following procedure: (a) starting from a specific form of ${\bf E}({\bf x})$ and ${\bf B}({\bf x})$, we first study the single particle motion, determining the constants of motions. (b) We look for a particular combination of such constants of motions that represents a DF, whose moments have a form close to the macroscopic structure we want to obtain. In particular, we require that the bulk velocity reproduces a planar shearing flow of the form ${\bf u}(x)=u(x) {\bf e}_u$, where ${\bf e}_u$ is a constant unit vector and the $x$ axis is perpendicular to ${\bf e}_u$. (c) Finally, calculating the moments of the DF we verify a posteriori that the remaining equations (\ref{far-amp})-(\ref{adiab}) are actually verified by the time-independent electromagnetic fields assumed at point (a). This procedure is the same as that used in Paper I for configurations different from those considered here. 

\section{Single-particle motion}\label{section3}
We define a Cartesian reference frame $S=\left\{x,y,z\right\}$, where $x$ is the direction of the electric field while the magnetic field is parallel to $yz$ plane: 
\begin{equation}\label{EB}
{\bf E}=E(x) {\bf e}_x \;\; ; \;\; {\bf B} = B_0  \left\{ \sin \left[ \Phi (x)\right] {\bf e}_y + \cos \left[ \Phi (x)\right] {\bf e}_z \right\}
\end{equation}
with $B_0$ the uniform magnetic field intensity, $\Phi(x)$ the local angle between ${\bf B}$ and the $z$-axis, and ${\bf e}_x$, ${\bf e}_y$ and ${\bf e}_z$ unit vectors along the Cartesian axes. In our solution the bulk velocity ${\bf u}={\bf u}(x)$ will be approximately directed along the y-axis. The two cases studied in Paper I correspond to $\Phi (x)= {\rm const}=0$ and $\Phi (x)={\rm const}= \pi/2$, $E(x)=0$. The electric field ${\bf E}$ orthogonal to ${\bf B}$ produces a drift in the particle motion that is partially responsible for the bulk velocity ${\bf u}$. 
In the particular case when $E(x)={\rm const}$ and $\Phi (x)={\rm const}$ each particle drifts with a uniform drift velocity given by ${\bf v}_d = c {\bf E} \times {\bf B}/B^2$. Therefore, in regions where the bulk velocity is uniform (outside shear layers) the electric and magnetic field are both uniform.
Finally, we observe that ${\bf j} \times {\bf B}=0$, where ${\bf j}$ is calculated through the Ampere's law (\ref{far-amp}). Therefore, in the considered configuration no macroscopic Lorentz force is exerted on the plasma by the magnetic field.

The single-particle Lagrangian is given by
\begin{equation}\label{lagrang}
\mathcal{L}(x,{\bf v},t)=\frac{m}{2} \left( v_x^2 + v_y^2 + v_z^2 \right) - e\phi(x;x_0) + \frac{e B_0}{c} \left[{\bf a}(x) \cdot {\bf v}\right]
\end{equation}
where $\phi(x;x_0) = -\int_{x_0}^x E(x') dx'$ is the electric potential, $x_0$ is a position where $\phi=0$, and ${\bf a}=a_y(x) {\bf e}_y + a_z(x) {\bf e}_z$ is a re-normalized vector potential: ${\bf B}=\nabla \times (B_0 {\bf a})$. Since $y$ and $z$ are cyclic coordinates, two constant of motions are given by:
\begin{equation}\label{wywz}
w_y = v_y - \Omega_p a_y(x) \;\; ; \;\; w_z = v_z - \Omega_p a_z(x)
\end{equation}
where $\Omega_p=e B_0/(mc)$ is the proton Larmor frequency. For given values of $w_y$ and $w_z$, relations (\ref{wywz}) univocally express the velocity components $v_y$ and $v_z$ as functions of the particle position $x$. Writing the Lagrange equation $d(\partial \mathcal{L}/\partial v_x)/dt - (\partial \mathcal{L}/\partial x) =0$ and using the relations (\ref{wywz}) to eliminate $v_y$ and $v_z$, we obtain an equation for the particle motion along $x$:
\begin{eqnarray}\label{1Dmot}
\frac{d v_x}{dt} = - \frac{d}{dx} \left\{ \frac{e}{m} \phi(x;x_0) - \Omega_p \left[ w_y a_y(x) + w_z a_z(x) \right]\right.\nonumber\\
\left. + \frac{\Omega_p^2}{2}\left[a_y^2(x)+a_z^2(x)\right]\right\}
\end{eqnarray}
The LHS of eq. (\ref{1Dmot}) can be re-written in the form $d(v_x^2/2)/dx$ and the equation can be integrated in the interval $\left[x_0,x\right]$. After some algebra, we obtain:
\begin{equation}\label{encons}
\frac{1}{2}m v_x^2 + U_{\rm eff}(x;x_0) = e_0
\end{equation}
where
\begin{eqnarray}\label{Ueff}
&& U_{\rm eff}(x;x_0) = e \phi(x;x_0) + \frac{1}{2} m \Omega_p^2 \left[ a_y(x) - a_y(x_0)\right]^2 \nonumber \\ 
&& + \frac{1}{2} m \Omega_p^2 \left[ a_z(x) - a_z(x_0)\right]^2 -m\Omega_p v_{0y} \left[ a_y(x) - a_y(x_0)\right] \nonumber \\
&& - m\Omega_p v_{0z} \left[ a_z(x) - a_z(x_0)\right]
\end{eqnarray}
with $e_0 = m v_{x0}^2/2$ a constant and $v_{0i}=v_i(x=x_0), i=x,y,z$. Equation (\ref{encons}) expresses the energy conservation for a
particle with mass $m$ following a 1D motion in the effective potential energy $U_{\rm eff}(x;x_0)$. In deriving the form (\ref{Ueff}) we have used eq. (\ref{wywz}) calculated at $x=x_0$ to express the constants $w_y$ and $w_x$ in terms of $v_{0y}$ and $v_{0z}$, respectively.

We want to describe a configuration where a finite number of shear layers is present, while far from that region the bulk velocity is uniform, corresponding to uniform ${\bf E}$ and ${\bf B}$. Therefore, for sufficiently large $|x|$ the scalar potential $\phi$ and the components $a_y$ and $a_z$ of the vector potential depend linearly on $x$: $\phi(x;x_0) \sim -E_{\pm \infty}x + {\rm const}$; $a_y(x) \sim B_{z,\pm \infty}x/B_0+ {\rm const}$; $a_z(x) \sim -B_{y,\pm \infty}x/B_0+ {\rm const}$, with $E_{\pm \infty}$, $B_{y,\pm \infty}$ and $B_{z,\pm \infty}$ asymptotic values. Inspecting expression (\ref{Ueff}), it is clear that for sufficiently large $|x|$ the dominant terms in $U_{\rm eff}(x;x_0)$ are proportional to $x^2$. Hence, the particle motion along $x$ is confined within a potential well: $x_m \le x \le x_M$, where $U_{\rm eff}(x_m;x_0) = U_{\rm eff}(x_M;x_0)=e_0$ and the particle moves back and forth in the interval $\left[x_m,x_M \right]$, with vanishing $v_x$ at $x_m$ and $x_M$. In other words, $x(t)$ and $v_x(t)$ are periodic function with a given period $\tau$, and then Eq.s (\ref{wywz}) imply that $v_y(t)$ and $v_z(t)$ are periodic with the same period $\tau$, too. Therefore, the particle follows a closed trajectory in the velocity space. In contrast, $y(t)$ and $z(t)$ are not necessarily periodic functions and the particle trajectory in the physical space is, in general, an open curve. We notice that such results are valid regardless of the specific field dependence on $x$, provided that the width of shear region is limited.

In consequence of periodicity, the time-averaged velocity over the period $\tau$ provides the drift velocity in the particle motion. Therefore, we define the guiding center $x$-position $x_c$ and velocity ${\bf v}_c$ as
\begin{equation}\label{xcvc}
x_c = \langle x \rangle_c = \frac{1}{\tau} \int_0^\tau x(t)dt \;\; ; \;\;   {\bf v}_c = \langle {\bf v} \rangle_\tau = \frac{1}{\tau} \int_0^\tau {\bf v}(t) dt
\end{equation}
Periodicity of $x(t)$ implies $\langle v_x\rangle_\tau =0$; hence, the guiding center velocity has only two components ${\bf v}_c = v_{cy}{\bf e}_y+v_{cz}{\bf e}_z = \langle v_y \rangle_\tau {\bf e}_y + \langle v_z \rangle_\tau {\bf e}_z$. Of course, $x_c$, $v_{cy}$ and $v_{cz}$ are constants of motion. In the particular case of uniform ${\bf B}$ (as in Paper I or in the next section), $a_y(x)$ and $a_z(x)$ are linear functions of $x$; as a consequence, the guiding center velocity ${\bf v}_c$ coincides with the projection of particle velocity onto the $yz$ plane, calculated at the guiding center position $x=x_c$. However, this latter property does not hold for a nonuniform ${\bf B}$. Finally, we notice that our definition (\ref{xcvc}) of guiding center is different from those used by \citet{ganguli88} and by \citet{cai90}. In particular, the definition by \citet{cai90} implies that a particle can have more than one guiding center, while in our approach a single guiding center is defined for each particle (see Paper I for a discussion). 

Another constant of motion is given by the total energy $\mathcal{E}$, whose value depend on the choice of location $x_0$ where the electric potential $\phi(x;x_0)$ vanishes. Though the value of $x_0$ does not affect the single particle motion, it is natural to choose $x_0=x_c$. In the case of a uniform electric and magnetic fields such a choice is in accordance with the macroscopic invariance of the fluid properties with $x$ (see Paper I). Therefore, the total energy is defined by
\begin{equation}\label{Etot}
\mathcal{E}=K+e\phi(x;x_c)=\frac{1}{2} m \left( v_x^2 + v_y^2 + v_z^2 \right) + e\phi(x;x_c) 
\end{equation}
where $K$ is the kinetic energy. 


\section{Uniform oblique ${\mathbf B}$}\label{section4}
We consider first a configuration corresponding to a uniform ${\bf B}$ obliquely oriented with respect to ${\bf u}(x)$. This kind of configuration has been considered in fluid models of KHI \citep[e.g.,][]{miura82} with applications to the magnetopause \citep{miura90}. Indeed, the oblique-${\bf B}$ configuration is more realistic for the magnetopause than a strictly perpendicular case \citep[e.g.,][]{settino20}. In the HVM framework an approximately stationary state has been obtained for this configuration by \citet{cerri13}. Here we will derive an exact stationary state. 

\begin{figure}
\resizebox{\hsize}{!}{\includegraphics{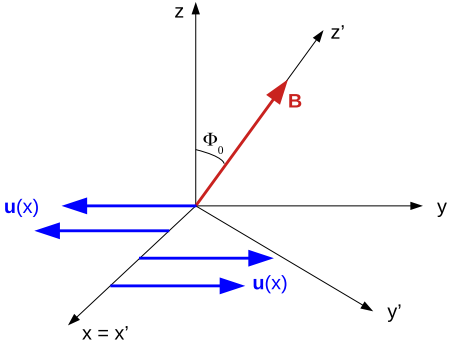}}
\caption{The reference frames $S=\left\{ x,y,z\right\}$ and $S'=\left\{ x',y',z'\right\}$ are represented and the orientation of ${\bf B}$ and ${\bf u}$ are indicated for the uniform ${\bf B}$ configuration.
\label{fig:frames}}
\end{figure} 

The present case can be described by Eq.s (\ref{EB}), when $\Phi(x)=\Phi_0={\rm const}$. We want to obtain a shearing flow where the bulk velocity ${\bf u}(x)$ is approximately directed along the $y$ direction, and therefore forms an angle $\theta=\pi/2 - \Phi_0$ with the magnetic field. The present situation reduces to the ``perpendicular" case studied in Paper I when $\Phi_0=0$. In order to exploit this analogy, it is useful to introduce a rotated reference frame $S'=\left\{ x',y',z'\right\}$, where $x'$ and $x$ axes coincide and $z'$ axis is parallel to ${\bf B}$. The two reference frames and other quantities are illustrated in Fig. \ref{fig:frames}. 
In this configuration no force is acting along $z'$ and therefore the particle velocity component $v_{z'}$ is a constant of motion. As a consequence, the ``perpendicular" energy $\mathcal{E}_\perp = \mathcal{E} - m v_{z'}^2/2$ is a constant of motion, as well.

Let us consider first the DF of the perpendicular case of Paper I [eq. (31) of Paper I] corresponding to the case $\Phi_0=0$:
\begin{equation}\label{fperp}
f_\perp(x,v_x,v_y,v_z) = C \exp \left[ -\frac{\mathcal{E}_0(x,v_x,v_y,v_z)}{m v_{th,p}^2} \right]
\end{equation}
where $\mathcal{E}_0=\mathcal{E}-m v_{c\perp}^2/2$ is the "reduced" total energy where the kinetic energy associated with the guiding center motion has been subtracted, and $v_{c\perp}$ is the guiding center velocity component perpendicular to ${\bf B}$ (the only nonvanishing component in this case).
$C$ is a normalization constant and $v_{th,p}$ is the proton thermal speed away from the shear region, where $f_\perp$ reduces to a SM.  
The associated bulk velocity ${\bf u}_\perp$ is also directed along the $y'$ direction (Paper I). 

In the oblique case ($\Phi_0\ne 0$) , we want to obtain a bulk velocity ${\bf u}$ in the $y$ direction, which has two components: ${\bf u}_\perp$ and ${\bf u}_{||}$, respectively perpendicular and parallel to ${\bf B}$. As in Paper I, the former component is due to the guiding center velocity $v_{c\perp}$ of particles. To obtain also a parallel bulk velocity ${\bf u}_{||}$ we add a shift $U_{||}$ of the DF in the velocity space along the $v_{z'}$ direction, similar to the shift that transforms a Maxwellian into a SM. This is coherent with the cyclicity of the $z'$ coordinate. 
The amount $U_{||}$ of the shift should (roughly) correspond to guiding centers moving along the $y$ direction. The geometry suggests the relation $U_{||}/v_{c\perp}= \tan \Phi_0$. On the base of the above considerations, we modify the expression (\ref{fperp}) and write the following form for the DF of uniform oblique ${\bf B}$:
\begin{eqnarray}\label{fC1}
f_{\Phi_0}(x,{\bf v}) &=& C \exp \left\{ -\frac{1}{2v_{th,p}^2} \left[ v_x^2 + \left[v_{y'}(v_y,v_z,\Phi_0)\right]^2\right. \right. \nonumber \\
& & \left. \left. + \left[ v_{z'}(v_y,v_z,\Phi_0) - v_{c\perp}(x,{\bf v}) \tan\Phi_0\right]^2 \right. \right. \nonumber \\
& & \left. \left. + e  \phi(x;x_c(x,{\bf v}))- \left[v_{c,\perp}(x,{\bf v})\right]^2
 \right]  \right\}
\end{eqnarray}
where $ v_{y'}(v_y,v_z,\Phi_0)=v_y\cos\Phi_0-v_z\sin \Phi_0$, $v_{z'}(v_y,v_z,\Phi_0)=v_y \sin\Phi_0 + v_z \cos\Phi_0$, and $v_{c,\perp}(x,{\bf v})$ is the guiding center velocity of a particle that at a given time is at position $x$ with velocity ${\bf v}$ (the knowledge of $x$ and ${\bf v}$ fully determines the particle motion). We observe that the form (\ref{fC1}) for $\Phi_0=0$ reduces to the distribution function obtained in the case ``perpendicular ${\bf B}$" [equation (31) of Paper I]. It is important to notice that the DF defined by Eq. (\ref{fC1}) is expressed only in terms of constants of motion. In fact, the terms $v_x^2+v_{y'}^2+e\phi(x;x_c)=2 \mathcal{E}_\perp/m$, $v_{z'}$, and $v_{c\perp}$ are all constants of motions. Therefore, the DF (\ref{fC1}) is a stationary solution of the Vlasov equation, provided that ${\bf E}$ and ${\bf B}$ remain stationary. This latter condition will be verified subsequently. 

The form (\ref{fC1}) has been heuristically induced, trying to generalize a case considered in Paper I. Therefore, the associated bulk velocity ${\bf u}$ must be calculated {\em a posteriori}, to verify to what extent it has the expected form.
We also notice that the form (\ref{fC1}) only implicitly defines the DF, because explicit expressions for the guiding center velocity $v_{c\perp}(x,{\bf v})$ and position $x_c(x,{\bf v})$ are not known for an arbitrary electric field profile $E(x)$. Explicit analytical results will be obtained in the particular case of a linear profile for the electric field, denoted as ``local approximation". In the general case a numerical method will be applied to calculate the DF and its moments.

\subsection{Local approximation}
In the particular case of an electric field with a linear profile the single particle motion can be analytically calculated. In this case, the form (\ref{fC1}) of the DF and of its moments can be explicitly written. In the following we briefly describe the main results of this case, while more details are given in Appendix \ref{App:A}. The electric field has the form:
\begin{equation}\label{Elin}
E(x) = E_0 + \alpha_0 (x-x_0)
\end{equation}
with $E_0$ and $\alpha_0$ constant. Equation (\ref{Elin}) can also be intepreted as a local first-order approximation of a general profile if $|x-x_0| \ll |E_0/\alpha_0|$. For $\alpha_0 \ge m\Omega_p^2/e$ particles moves along $x$ linearly or exponentially in time, leading to a breakdown of the local approximation. Therefore, we consider only the case $\alpha_0 < m\Omega_p^2/e$, when the particle orbit in the velocity space is given by a shifted ellipse located in a plane parallel to the $v_{x'}v_{y'}$ plane, which is travelled with a gyrofrequency $\omega=(\Omega_p^2 - e\alpha_0/m)^{1/2}$. The guiding center is the ellipse center (the explicit expression of $x_c(x,{\bf v})$ is given in Appendix \ref{App:A}); it moves with a transverse velocity that coincides with the ${\bf E}\times {\bf B}$ drift velocity calculated at the guiding center position: $v_{c\perp}(x,{\bf v})=-cE(x_c)/B$ . Using the expressions of $x_c(x,{\bf v})$ and $v_{c\perp}(x,{\bf v})$, the explicit form of the DF can be derived:
\begin{eqnarray}\label{fC1la}
& & f_{\Phi_0}^{(la)}(x,{\bf v})=\frac{n_0}{(2\pi)^{3/2} v_{th,p}^3} \sqrt{\frac{\Omega_p B_0}{\Omega_p B_0-c\alpha_0}} \exp \left\{-\frac{1}{2v_{th,p}^2} \right. \nonumber \\
& & \left. \times \left\{ v_x^2 + 
\frac{\Omega_p B_0}{\Omega_p B_0-c\alpha_0} \left[ \frac{c\alpha_0}{B_0} \left(x-x_0\right)+v_{y'}-v_{d0}\right]^2 \right.\right. \nonumber \\
& & \left.\left. + \left[ v_{z'} - v_{d0} \tan\Phi_0 \right.\right.\right.\nonumber \\ 
& & \left.\left.\left. + \frac{c\alpha_0 \tan\Phi_0}{\Omega_pB_0-c\alpha_0} \left[ \Omega_p\left(x-x_0\right) + v_{y'} - v_{d0}\right]\right]^2\right\}\right\}
\end{eqnarray}
where $n=n_0$ is the uniform density and $v_{d0}=-cE_0/B_0$ is the drift velocity at the position $x_0$.  The associated bulk velocity has the form:
\begin{equation}\label{uC1loc}
{\bf u(x)} = -\frac{cE(x)}{B_0} {\bf e}_{y'} - \frac{cE(x)}{B_0} \tan \Phi_0 {\bf e}_{z'} = - \frac{cE(x)}{B_0\cos \Phi_0} {\bf e}_y
\end{equation}
The expression (\ref{uC1loc}) corresponds to a planar shearing flow varying along $x$ and directed at constant angle $\theta=\pi/2 - \Phi_0$ with ${\bf B}$, at each $x$. Therefore, in the local approximation the DF (\ref{fC1}) exactly reproduces the desired form for the bulk velocity. In particular, the bulk velocity component $u_\perp=u_{y'}$ perpendicular to ${\bf B}$ is equal to the local ${\bf E}\times {\bf B}$ drift velocity.

\subsection{General electric field profile}
In the case of electric field with a general profile $E(x)$ we have used a numerical procedure to calculate the DF (\ref{fC1}) and its moments. In the procedure, where the motion equations of single particles are integrated, the periodicity in the velocity space represents a key element. Of course, constants of motions appearing in the expression (\ref{fC1}) can be evaluated at any instant of time; therefore, the phase-space coordinates $\left\{ x, {\bf v} \right\}$ are interpreted as the particle position and velocity at the initial time of its motion. In the numerical procedure, such initial conditions are taken on a regular grid in the phase-space (1D in the physical space and 3D in the velocity space): $x(t=0)=x_i$ , $v_x(t=0)=v_{x,j}$, $v_y(t=0)=v_{y,k}$, $v_z(t=0)=v_{z,l}$ ($i,j,k,l$ are indexes which span on the 4D grid and identify the given particle). Starting from this set of initial conditions, the motion equations of each particle are integrated during one single period $\tau_{ijkl}$. For the time integration we used a 3rd-order Adams-Bashforth scheme. Then, constants of motion are calculated: the guiding center transverse velocity $v_{c\perp,ijkl}=\langle v_{\perp} \rangle_{\tau_{ijkl}}$ and position $x_{c,ijkl}=\langle x \rangle_{\tau_{ijkl}}$, as well as the transverse energy $U_{\perp,ijkl}=m(v_{y,k}^2+v_{z,l}^2)/2-\int_{x_{c,ijkl}}^{x_i} E(x)dx$. Finally, these quantities, along with the parallel velocity $v_{z',kl}=v_{y,k}\cos\Phi_0 - v_{z,l}\sin\Phi_0$, are inserted into the expression (\ref{fC1}) and the value $f_{\phi_0}(x_i,v_{x,j},v_{y,k},v_{z,l})$ of the DF at the given phase-space point is calculated.

The set of values of the DF calculated on the grid points allow to obtain the moments by numerically calculating the corresponding integrals. In particular, we found that the $x$ component of the resulting bulk velocity is vanishing: $u_x=0$, similar as in the local approximation [Eq. (\ref{uC1loc})]. Therefore, the term ${\bf u}\times {\bf B}$ in the generalized Ohm's law (\ref{ohm}) is directed in the $x$ direction. Moreover, in this configuration it is ${\bf j}=0$. Hence, only the $x$ component of equation (\ref{ohm}) is nonvanishing, which implies:
\begin{equation}\label{dpedx} 
\frac{dp_e}{dx} = -e n(x) \left\{ E(x) + \frac{B_0}{c}\left[ u_y(x)\cos \Phi_0 - u_z(x) \sin \Phi_0 \right]\right\}
\end{equation}
This equation determines the electron pressure $p_e(x)$. 

Concerning the time dependence, let us assume that the obtained configuration holds at the initial time. Since $\partial f_{\Phi_0}/\partial t=0$, all the moments of $f_{\Phi_0}$ have a vanishing time derivative; in particular, it is $\partial n/\partial t=0$. Being $\nabla \times {\bf E}=0$, the Faraday's law (\ref{far-amp}) implies that $\partial {\bf B}/\partial t =0$. Moreover, being $({\bf u}\cdot \nabla)(p_e/n^{\gamma_e})=0$, Eq. (\ref{adiab}) implies $\partial p_e/\partial t=0$. Hence, from the Ohm's law (\ref{ohm}) it follows $\partial {\bf E}/\partial t=0$. We can conclude that the considered configuration is a stationary solution of the {\em full set} of HVM equations (\ref{vlasov})-(\ref{adiab}). 

\subsection{Results for a double shear configuration}

To illustrate results we consider a particular configuration with two opposite shear layers. We introduce rescaled variables: magnetic field is normalized to $B_0$, time to $\Omega_p^{-1}$, density to the value $\bar{n}$ away from the shear regions, velocities to the Alfv\'en velocity $v_A=B_0/(4\pi m \bar{n})^{1/2}$, lengths to the proton inertial length $d_p=v_A/\Omega_p$, electric field to $v_A B_0/c$, and electron pressure to $\bar{n}mv_A^2$. To simplify the notation, in the remainder of this subsection we will denote rescaled quantities with the same symbols as the original quantities. We choose a spatial domain $D$ defined by $0 \le x \le L$, with $L=100$ and the following form for the electric field:
\begin{equation}\label{EC1}
E(x) = E_0 \left[1 - \tanh\left(\frac{x-L/4}{\Delta x}\right) + \tanh\left(\frac{x-3L/4}{\Delta x}\right)\right]
\end{equation}
where $E_0=\cos\Phi_0$ and $\Delta x=2.5$. This corresponds to two opposite shear layers of width $\Delta x$ located at $x=L/4$ and $x=3L/4$, respectively. The electric field (\ref{EC1}) is periodic in the domain $D$: $E(0)=E(L)\simeq E_0$, while in the center of the domain it is $E(L/2)\simeq -E_0$. Moreover, $E(x)$ vanishes in the limit $\Phi_0 \rightarrow \pi/2$, consistent with the case when ${\bf B}$ is parallel to ${\bf u}$ (Paper I). 

We have calculated the DF (\ref{fC1}) using the above-described procedure on a grid formed by $N_x=256$ points in the $x$ direction and $N_v=71$ points in each velocity direction $v_i,\,i=x,y,z$, where $-7 v_{th,p} \le v_i \le 7 v_{th,p}$, with $v_{th,p}=1$ (in scaled units). The $N_x \, N_v^3 \simeq 9.16\times 10^7$ particle trajectories have been calculated using a parallel computing procedure. 

\begin{figure}
\resizebox{\hsize}{!}{\includegraphics{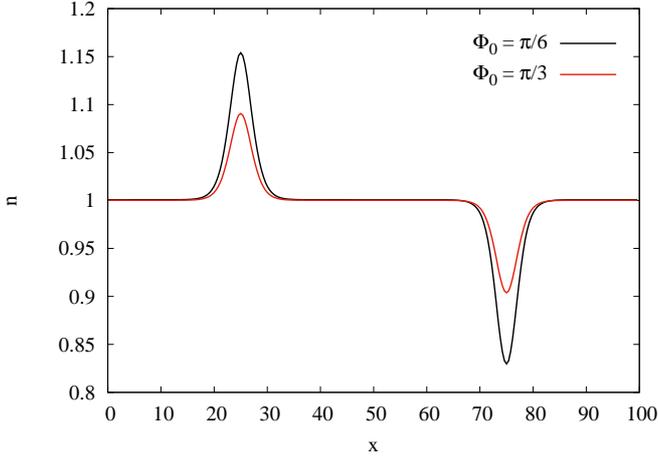}}
\caption{Profiles of density $n(x)$ calculated for $\Phi_0=\pi/6$ (black line) and $\Phi_0=\pi/3$ (red line), in the uniform oblique ${\bf B}$ case.
\label{fig:densC1}}
\end{figure} 

The associated density $n(x)=\int f_{\Phi_0}(x,{\bf v}) d^3{\bf v}$ is plotted in Fig. \ref{fig:densC1}, in the cases $\Phi_0=\pi/6$ and $\Phi_0=\pi/3$.
In contrast with the local approximation, now the density $n(x)$ is not uniform, except away from the two shear regions. The non-uniformity of $n(x)$ decreases with increasing $\Phi_0$; actually, when ${\bf B}$ is parallel to ${\bf u}$ (corresponding to $\Phi_0=\pi/2$) the density profile is uniform (Paper I). In particular, $n(x)$ is maximum (minimum) at center of the shear at $x=L/4$ ($x=3L/4$). This asymmetry between the two shear layers is a genuine kinetic effect, related to the sign of the dot product $(\nabla \times {\bf u})\cdot \mathbf{\Omega}_p$, ($\mathbf{\Omega}_p$ being the proton vector gyration angular velocity) which is opposite in the two shear layers. 

\begin{figure}
\resizebox{\hsize}{!}{\includegraphics{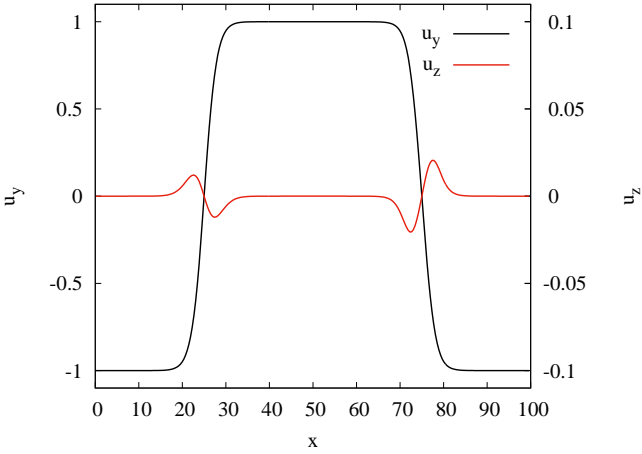}}
\caption{Profiles of bulk velocity components $u_y(x)$ (black line) and $u_z(x)$ (red line) calculated for $\Phi_0=\pi/6$, in the uniform oblique ${\bf B}$ case.
\label{fig:velC1}}
\end{figure} 

The profiles of the bulk velocity components $u_i(x)=\int v_i f_{\Phi_0}(x,{\bf v}) d^3{\bf v}/n(x)$, $i=y,z$ are plotted in Fig. \ref{fig:velC1} for $\Phi_0=\pi/6$. We notice that ${\bf u}$ is not exactly in the $y$ direction, as in the local approximation case. However, $|u_z|$ is about two orders of magnitude smaller than $|u_y|$, while $|u_x|\sim 10^{-12}$ (not shown). Therefore, the condition that ${\bf u}(x)$ and ${\bf B}$ form a constant angle $\theta=(\pi/2-\Phi_0)$ is well satisfied. The largest deviations from this condition are localized at the shear layers.

Moreover, the profile of $u_z(x)$ is qualitatively similar to $-d^2 E(x)/dx^2$; this is not surprising, since for a linear electric field profile (\ref{Elin}) it is $u_z(x)=0$.

\begin{figure}
\centering
\begin{minipage}[ht]{\linewidth}
   \centering
\includegraphics[width=\textwidth]{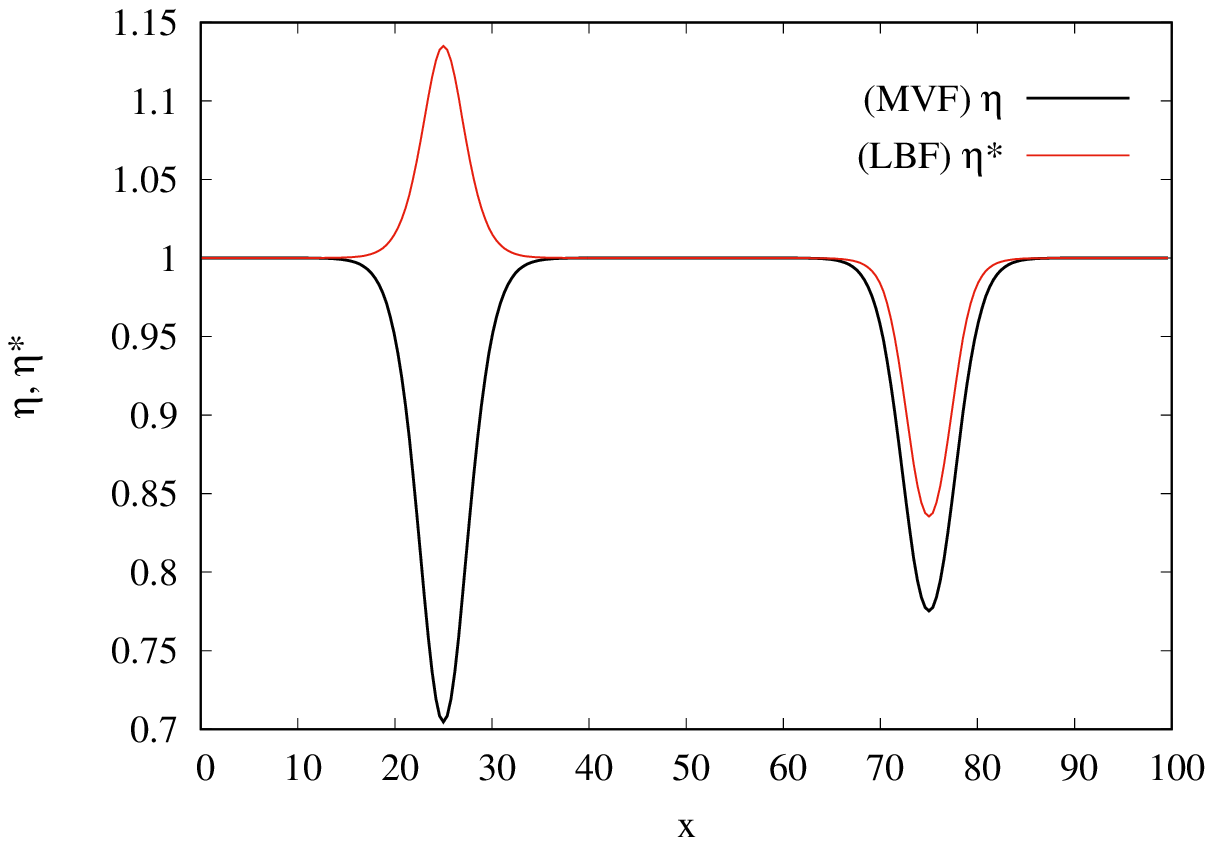}
\end{minipage}
\begin{minipage}[ht]{\linewidth}
   \centering
\includegraphics[width=\textwidth]{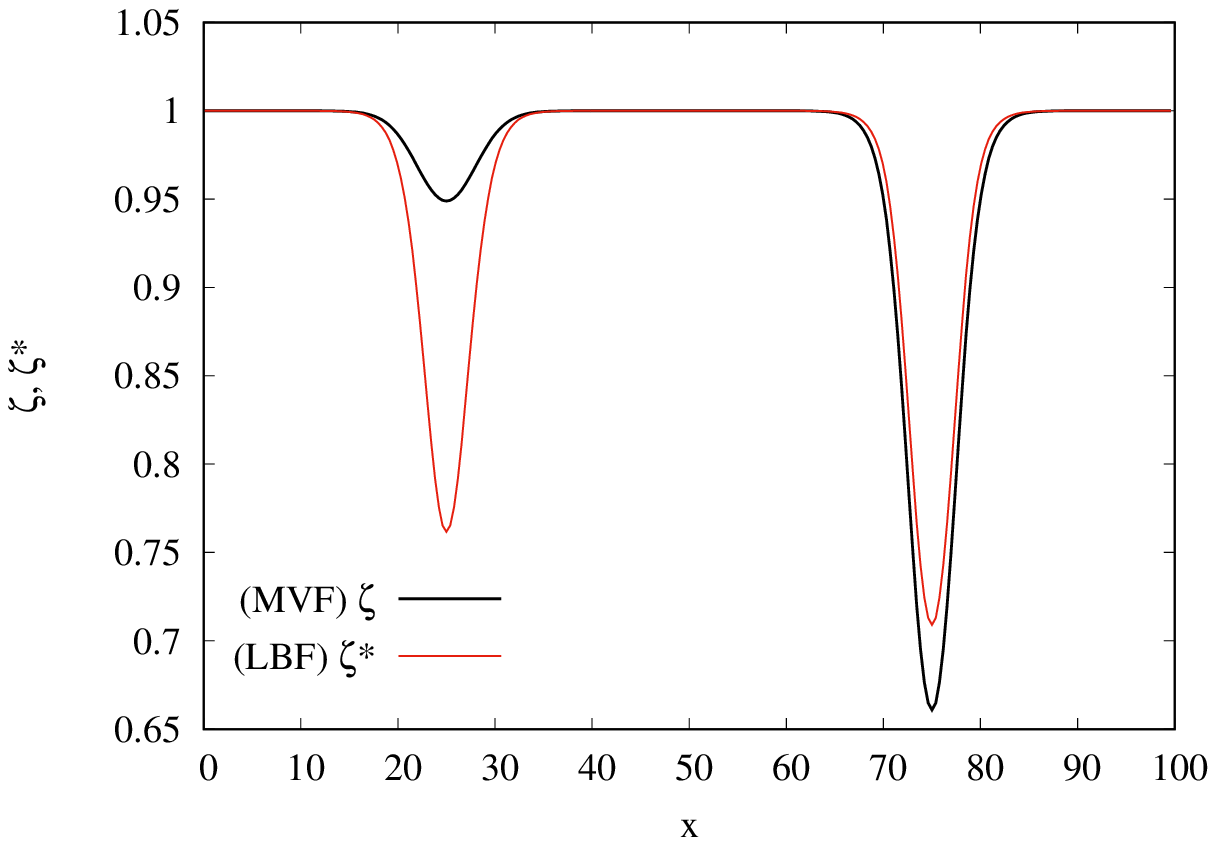}
\end{minipage}
\caption{Upper panel: profiles of anisotropy parameters $\eta$ (black line) and $\eta^*$ (red line); lower panel: profiles of agyrotropy parameters $\zeta$ (black line) and $\zeta^*$ (red line). Both panels refer to $\Phi_0=\pi/6$ in the uniform oblique ${\bf B}$ case.
\label{fig:anisagirC1}}
\end{figure}

In the shear regions the DF deviates from a Maxwellian. In particular, a temperature anisotropy develops that can be quantitatively evaluated by calculating the variance matrix. This has been done in the reference frame $S'$ where ${\bf B}$ is along the $z'$ axis. We refer to $S'$ as ``local ${\bf B}$ frame" (LBF). The corresponding variance matrix is
\begin{equation}\label{varmatr}
\sigma_{ij}^{LBF}(x)=\frac{1}{n(x)} \int \left[ v_i-u_i(x)\right] \left[ v_j-u_j(x)\right] f_{\Phi_0}(x,{\bf v}) d^3{\bf v}
\end{equation}
with $i,j=x',y',z'$. Normalized parallel and perpendicular temperatures are given by $T_{||}=\sigma_{z'z'}^{LBF}$ and $T_\perp=(\sigma_{x'x'}+\sigma_{y'y'})/2$, respectively, while the anisotropy and agirotropy indexes are $\eta^*=T_\perp/T_{||}$ and $\zeta^*=\min(\sigma_{x'x'}^{LBF},\sigma_{y'y'}^{LBF})/\max(\sigma_{x'x'}^{LBF},\sigma_{y'y'}^{LBF})$, respectively (a gyrotropic DF corresponds to $\zeta^*=1$, otherwise it is $\zeta^*<1$). Since the variance matrix is symmetric, a reference frame exists (denoted as minimum variance frame, MVF) where it is diagonal. Therefore, the eigenvalues $\lambda^{(3)} < \lambda^{(2)} < \lambda^{(1)}$ of $\sigma_{ij}$ give the temperatures along the MVF axes. The corresponding anisotropy and agyrotropy indexes are defined by $\eta=(\lambda^{(2)}+\lambda^{(3)})/(2\lambda^{(1)})$ and $\zeta=\lambda^{(3)}/\lambda^{(2)}$, respectively. In Fig. \ref{fig:anisagirC1} the profiles of $\eta$, $\eta^*$, $\zeta$ and $\zeta^*$ are plotted as functions of $x$, for $\Phi_0=\pi/6$. In the shear layers the DF is both anisotropic and agyrotropic. Anisotropy is larger in the MVF than in the LBF in both shear layers. Agyrotropy is larger in the LBF than in the MVF in the shear layer at $x=L/4$, the reverse holds at $x=3L/4$ , owing to a different orientation between $\Omega_p$ and $\nabla \times {\bf u}$ at each shear.

A similar behaviour holds also for $\phi_0=\pi/3$, except for a larger agyrotropy in the MVF in both shears (not shown).

\begin{figure}
\resizebox{\hsize}{!}{\includegraphics{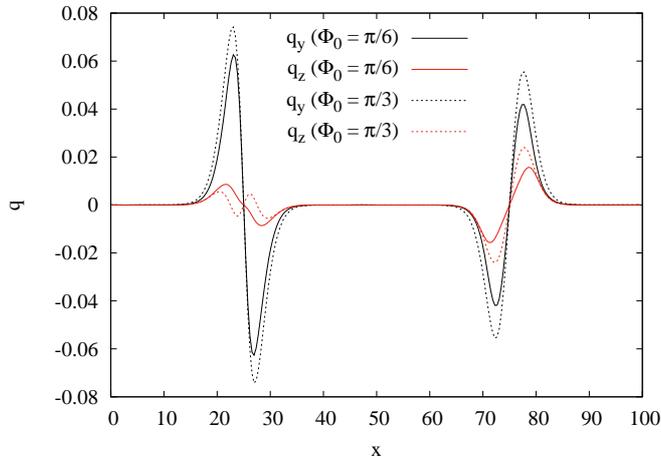}}
\caption{Profiles of heat flux components $q_y$ (black lines) and $q_z$ (red lines) for $\Phi_0=\pi/6$ (full lines) and $\Phi_0=\pi/3$ (dashed lines), in the uniform oblique ${\bf B}$ case. 
 \label{fig:calC1}}
\end{figure}

Finally, we calculated the heat flux ${\bf q}$, defined by
\begin{equation}\label{q}
{\bf q}(x)=\frac{1}{2} \int \left[ {\bf v}-{\bf u}(x)\right] |\left[ {\bf v}-{\bf u}(x)\right]|^2 f_{\Phi_0}(x,{\bf v}) d^3{\bf v}
\end{equation}
The components $q_y$ and $q_z$ are plotted in Fig. \ref{fig:calC1} as functions of $x$, for $\Phi_0=\pi/6$ and $\Phi_0=\pi/3$, while $q_x$ is null for all $x$ values. We see that $q_y$ and $q_z$ are not vanishing in the shear layers, where they have opposite behaviours. However, the average of both $q_y$ and $q_z$ across each shear layer is null.  

\subsection{Numerical simulation}
Though the above-described configuration is in principle exactly stationary, being derived through a numerical procedure it is necessarily affected by errors that arise during time integration of single particle motion. Moreover, in the perspective of applying such results in numerical simulations, other errors would appear related to discretization in the phase space and time that is inherent to numerical codes. Therefore, it is interesting to study the time behaviour of the above-described stationary state, when it is used as initial condition in a HVM numerical simulation. This has the purpose to find a quantitative evaluation of departure from stationarity due to numerical errors.

\begin{figure}
\resizebox{\hsize}{!}{\includegraphics{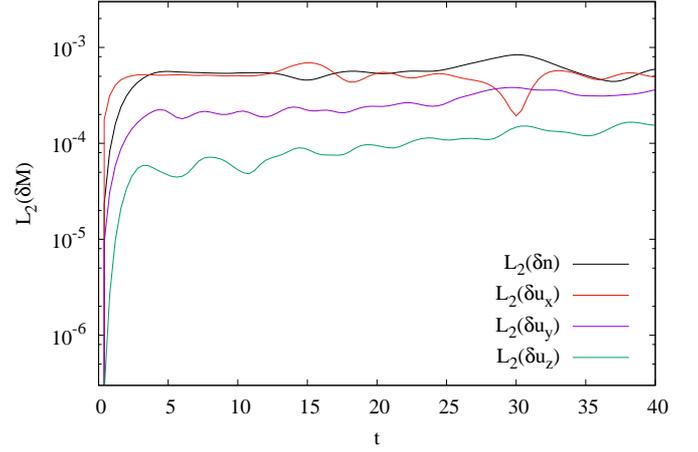}}
\caption{Profiles of $\L_2(\delta n)$ (black line), $\L_2(\delta u_x)$ (red line), $\L_2(\delta u_y)$ (purple line), and $\L_2(\delta u_z)$ (green line) as functions of time, as they result from the HVM simulation.
\label{fig:simulC1}}
\end{figure}

A simulation has been performed using a HVM numerical code, where periodic boundary conditions are applied in the physical space while it is assumed that the DF vanishes at the boundaries of the domain in the velocity space ($|v_{x,y,z}|\ge v_{max}=7v_{th,p}$). Details on the code can be found in \citet{valentini07}. The simulation lasts up to the time t=40 in normalized units. We calculated the time evolution of density and bulk velocity components, evaluating the departures from their initial profiles. In particular, for a given moment $M(x,t)$ we calculated the corresponding $L_2$-norm departure with respect to the initial profile $M(x,0)$:
\begin{equation}\label{L2M}
L_2(\delta M)(t)=\left\{ \frac{1}{L} \int_0^L \left[ M(x,t)-M(x,0) \right]^2 dx \right\}^{1/2}
\end{equation} 
In Fig.s \ref{fig:simulC1} the density $\L_2(\delta n)$ and bulk velocity $L_2(\delta u_i)$ ($i=x,y,z$) departures are plotted as functions of $t$, as they result from the simulation. As expected, such departures do not remain null. However, after a fast increase lasting few time units, both $L_2(\delta n)$ and $L_2(\delta u_i)$ saturate at a level that has an order of magnitude between $10^{-4}$ and $10^{-3}$. We conclude that numerical errors affect the stationarity of the solution to a low extent.

\section{${\bf B}$ with uniform intensity and variable direction.}\label{section5}

The second configuration, corresponds to a nonuniform magnetic field ${\bf B}$ whose direction changes in space, but keeping $|{\bf B}|$ uniform. This kind of configuration is more realistic for the magnetopause than the previous case where ${\bf B}$ had a constant direction. In fact, in many circumstances the magnetic field direction on the solar wind side is different from that on the magnetosphere side \citep[see, e.g.,][]{faganello17}. Moreover, if the magnetic field component parallel to the bulk velocity changes sign across the shear layer, magnetic reconnection can take place during the development of KHI, with consequences on the ion transfer from the solar wind to magnetosphere \citep{nakamura11,nakamura13}.
The present case corresponds to a configuration described by equations (\ref{EB}) where the angle $\Phi(x)$ between ${\bf B}(x)$ and the $z$-axis is not constant. In this case, particle trajectories are not planar curves. 

In order to have indications on how to find an appropriate stationary solution, we have first examined a local approximation.
We consider electric and magnetic field varying linearly with $x$, assuming the form (\ref{Elin}) for $E(x)$ and the equations
\begin{equation}\label{Bylin}
B_y(x)=B_0 \left[ b_{0y} + \gamma_y(x-x_0)\right]
\end{equation}
\begin{equation}\label{Bzlin}
B_z(x)=B_0 \left[ b_{0z} - \gamma_z(x-x_0)\right]
\end{equation}
where $b_{0y,0z}$ and $\gamma_{y,z}$ are constants and $|x-x_0| \ll |b_{0y,0z}/\gamma_{y,z}|$. In studying the particle dynamics, all terms that are quadratic or of higher order in the ratios $|\gamma_{y,z}(x-x_0)/b_{0y,0z}|$ will be neglected. Here we give only the main results of the procedure, while details can be found in the Appendix \ref{App:B}. The particle motion along $x$ corresponds to a harmonic oscillator $x(t)=R_0 \sin(\omega t + \varphi) + x_c$, where the $R_0$ is amplitude, $\varphi$ is the phase, the frequency is
\begin{equation}\label{omega}
\omega = \left[ \Omega_p^2 + \Omega_p \left( w_y\gamma_z + w_z\gamma_y\right) - \frac{e\alpha_0}{m} \right]^{1/2}
\end{equation}
and $x_c$ is the center of the oscillation, corresponding to the guiding center position (the expression of $x_c$ is given in Appendix \ref{App:B}). This solution holds if $\alpha_0 < (m/e)\left[ \Omega_p^2 + \Omega_p \left( w_y\gamma_z + w_z\gamma_y\right) \right]$, otherwise the local approximation breaks down. Coordinates $y(t)$ and $z(t)$ can be calculated using the equations (\ref{wywz}). Due to the quadratic form of $a_y(x)$ and $a_z(x)$, the coordinates $y(t)$ and $z(t)$, as well as the velocity components $v_y(t)$ and $v_z(t)$ are no longer harmonic functions of time. This implies that the closed trajectory in the velocity space is not an ellipse, as in the previous configuration and, in general, it is not planar. However, the following simple relation holds for the components $v_{cy}=\langle v_y \rangle$ and $v_{cz}=\langle v_z \rangle$ of the guiding center velocity: $v_{cz} B_y(x_c) - v_{cy} B_z(x_c) = cE(x_c)$, which implies that the guiding center velocity component perpendicular to ${\bf B}(x_c)$ is equal to the local ${\bf E}\times {\bf B}$ drift velocity:
\begin{equation}\label{vcperp}
v_{c\perp}=c \frac{{\bf E}(x_c) \times {\bf B}(x_c)}{B_0^2}
\end{equation}
This same property of the guiding center velocity was found in the oblique ${\bf B}$ configuration. This analogy will be exploited to generalize the DF found in the case of oblique uniform ${\bf B}$ to the present case of non-uniform ${\bf B}$. For this purpose, we go back to the uniform-${\bf B}$ case and re-write the argument of the exponential in Eq. (\ref{fC1}) (valid in the former case) in the following form: 
\begin{equation}\label{FC1}
F_{\Phi_0}=-\frac{2\mathcal{E}/m-v_{z'}^2+(v_{z'}-v_{c\perp}\tan\Phi_0)^2-v_{c\perp}^2}{2v_{th,p}^2}
\end{equation}
The velocity component $v_{z'}$ in this expression can be interpreted as the guiding center velocity component parallel to ${\bf B}$: $v_{z'}=v_{c||}$. Therefore, we write 
\begin{equation}\label{FC1a}
F_{\Phi_0}=-\frac{2\mathcal{E}/m-v_{c||}^2+\left[v_{c||}-U_{||}(v_{c\perp})\right]^2-v_{c\perp}^2}{2v_{th,p}^2}
\end{equation}
where $U_{||}(v_{c\perp})=v_{c\perp}\tan\Phi_0$. The expression (\ref{FC1a}) is valid in any reference frame and therefore it is suitable to be generalized to the nonuniform-${\bf B}$ case. Indeed, to obtain such a generalization we have only to specify another form for the function $U_{||}(v_{c\perp})$, that has the role of generating a bulk velocity component parallel to ${\bf B}$. In the case of nonuniform ${\bf B}$, the magnetic field changes direction with $x$ while the bulk velocity ${\bf u}$ must keep parallel to the $y$ axis for any $x$. Indicating by ${\bf e}_{||}={\bf B}/B$ and ${\bf e}_{\perp}={\bf e}_{||}\times {\bf e}_x$ the unit vectors parallel and perpendicular to ${\bf B}$, we can assume that ${\bf u} \sim U_{||} {\bf e}_{||} + v_{c\perp} {\bf e}_\perp$. Imposing $u_z=0$ in the previous expression we derive the relation $U_{||}=(B_y/B_z) v_{c\perp}=v_{c\perp} \tan\Phi $. On the base of the above considerations, we assume the following form for the DF in the general case:
\begin{eqnarray}\label{fC2}
f_{\Phi(x)}(x,{\bf v}) = C \exp \left\{ -\frac{1}{2v_{th,p}^2} \left\{ v_x^2 + v_{y}^2 + v_{z}^2 + \right.\right. \nonumber \\
\left.\left. e\phi(x;x_c(x,{\bf v}))- \left[v_{c||}(x,{\bf v})\right]^2 - \left[v_{c\perp}(x,{\bf v})\right]^2\right. \right. \nonumber \\
\left. \left. + \left[ v_{c||}(x,{\bf v}) - v_{c\perp}(x,{\bf v}) \tan\Phi(x_c(x,{\bf v})) \right]^2
 \right\}  \right\}
\end{eqnarray}
The associated electron pressure $p_e$ is calculated by the equation (\ref{adiab}). We will show that the bulk velocity component $u_x$ is vanishing; moreover, it is ${\bf j}\times {\bf B}=0$. Therefore, the only nonvanishing component of Eq. (\ref{adiab}) reduces to 
\begin{eqnarray}\label{dpedxC2}
\frac{dp_e}{dx} = -en(x)\left\{ E(x) + \frac{B_0}{c} \left[ u_y(x) \cos\Phi(x) \right.\right. \nonumber \\
\left.\left. - u_z(x) \sin\Phi(x) \right]\right\}
\end{eqnarray} that determines the profile $p_e(x)$. 

We observe that $f_{\Phi(x)}(x,{\bf v})$ is expressed only in terms of the constants of motion: total energy $\mathcal{E}$, guiding center position $x_c$ and velocity ${\bf v}_c$. Therefore, the DF (\ref{fC2}) is a stationary solution of the Vlasov equation (\ref{vlasov}), provided that the electric and magnetic field remain constant in time. Indeed, using the same arguments as in the case of uniform ${\bf B}$ it can be shown that $\partial {\bf E}/\partial t = \partial {\bf B}/\partial t = \partial p_e/\partial t =0$. We conclude that the considered configuration is a stationary solution of the whole set of equations (\ref{vlasov})-(\ref{adiab}). We also notice that the expression (\ref{fC2}) reduces to the form (\ref{fC1}) when $\Phi(x)=\Phi_0={\rm const}$.

\begin{figure}
\resizebox{\hsize}{!}{\includegraphics{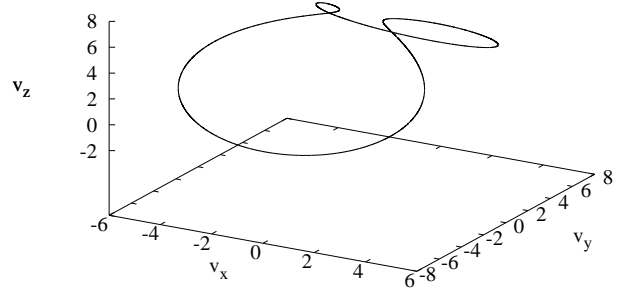}}
\caption{Trajectory in the velocity space of a particle starting from $x=75$, $ v_x=-1$, $v_y=0.5$, $v_z=6.5$.\label{fig:orbitC2}}
\end{figure} 

\subsection{Velocity and magnetic field shear}

For general profiles of $E(x)$ and $\Phi(x)$ an explicit evaluation of the DF (\ref{fC2}) can be obtained using the same numerical procedure as in the case $\Phi(x)=\Phi_0={\rm const}$, which determines $f_{\Phi(x)}$ on a 4D grid in the $\left\{ x, {\bf v} \right\}$ phase space. In particular, for the sake of illustration we consider a particular case where $E(x)$ is given by Eq. (\ref{EC1}) and 
\begin{equation}\label{phix}
\Phi(x)=\Delta\Phi \left[ \tanh \left( \frac{x-L/4}{\Delta x} \right)- \tanh \left( \frac{x-3L/4}{\Delta x} \right) -1\right]
\end{equation}
In this configuration, ${\bf B}$ rotates by an angle $\pm 2\Delta \Phi$ across the shear layers, while it is essentially uniform away from the shear layers. The $B_y$ component changes sign across the shear layers, while ${\bf B}$ is in the $z$ direction at the center of each shear layer, $x=L/4$ and $x=3L/4$, where $\Phi=0$. The width $\Delta x$ of the magnetic shears is the same as that of the velocity shears.

\begin{figure}
\resizebox{\hsize}{!}{\includegraphics{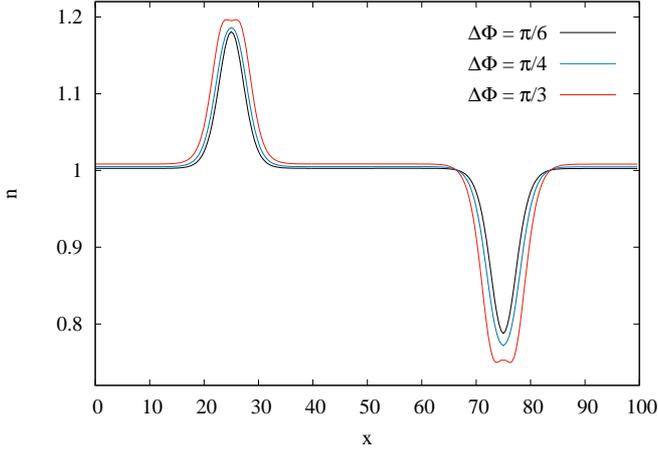}}
\caption{Profiles of density $n$ as a function of $x$ calculated for $\Delta \Phi=\pi/6$ (black line), $\Delta \Phi=\pi/4$ (blue line), and $\Delta \Phi=\pi/3$ (red line) in the case of nonuniform ${\bf B}$.
\label{fig:densC2}}
\end{figure} 

Numerical results are presented for $L=100$, $\Delta x=2.5$ (in normalized units) and for various values of $\Delta\Phi$. In Fig. \ref{fig:orbitC2} we plot the trajectory of a particle in the velocity space. The curve is more complex than those typically found in the case of uniform ${\bf B}$ and is not planar. We have calculated the moments of the DF (\ref{fC2}) for this configuration. In Fig. \ref{fig:densC2} the density profile $n(x)$ is plotted; as for the uniform-${\bf B}$ case, $n(x)$ has a local maximum or minimum at shear layers, according to the sign of $(\nabla \times {\bf u}) \cdot \mathbf{\Omega}_p$ at shear layers.

This asymmetry tends to increase and the profile $n(x)$ becomes more complex for increasing $\Delta\Phi$. For all the values of $\Delta\Phi$, the DF has been normalized by imposing that $\int_0^L n(x)dx/L=1$.

\begin{figure}
\resizebox{\hsize}{!}{\includegraphics{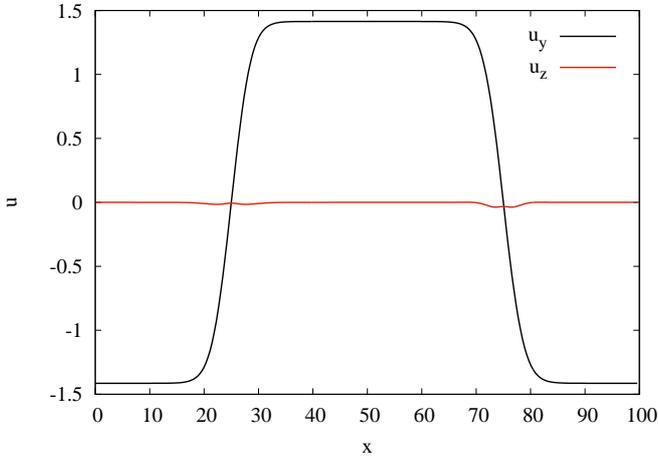}}
\caption{Profiles of bulk velocity components $u_y$ (black line) and $u_z$ (red line) as functions of $x$ calculated for $\Delta \Phi=\pi/4$ in the case of nonuniform ${\bf B}$.
\label{fig:velC2}}
\end{figure} 

\begin{figure}
\centering
\begin{minipage}[ht]{\linewidth}
   \centering
\includegraphics[width=\textwidth]{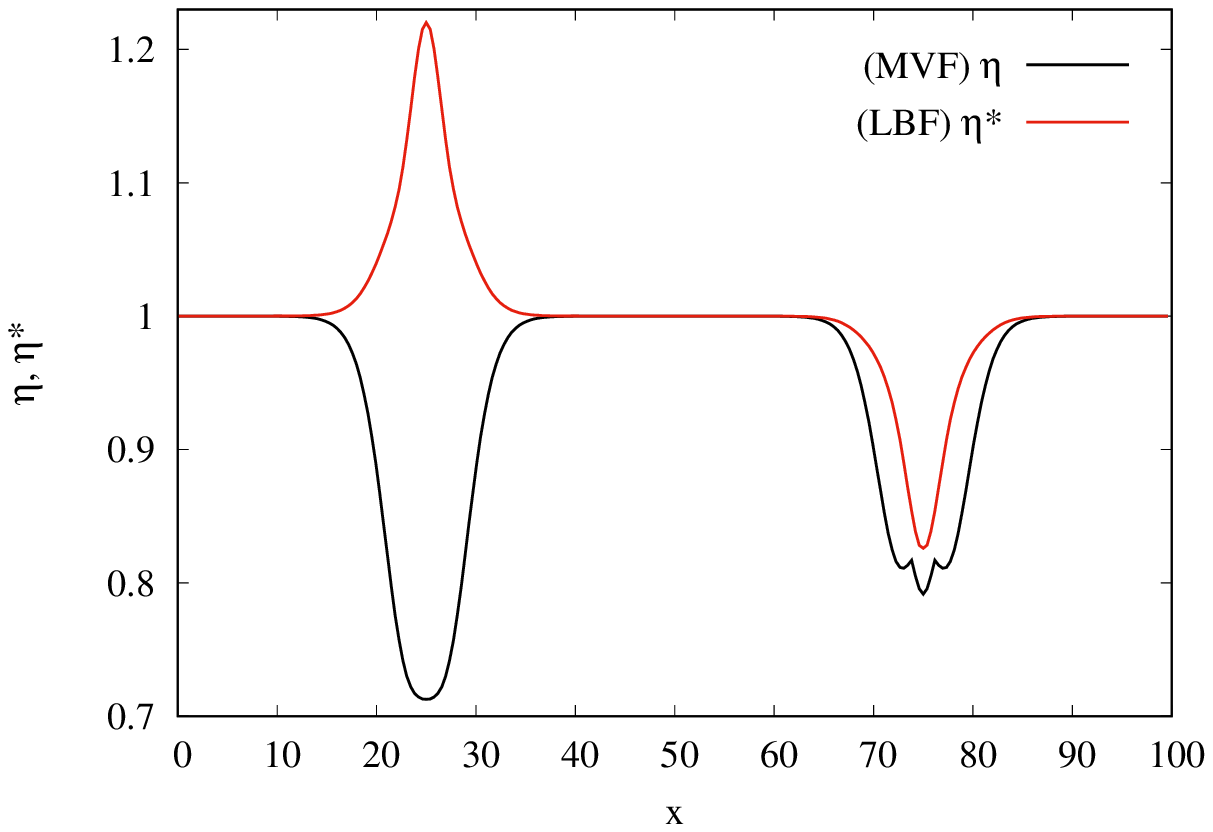}
\end{minipage}
\begin{minipage}[ht]{\linewidth}
   \centering
\includegraphics[width=\textwidth]{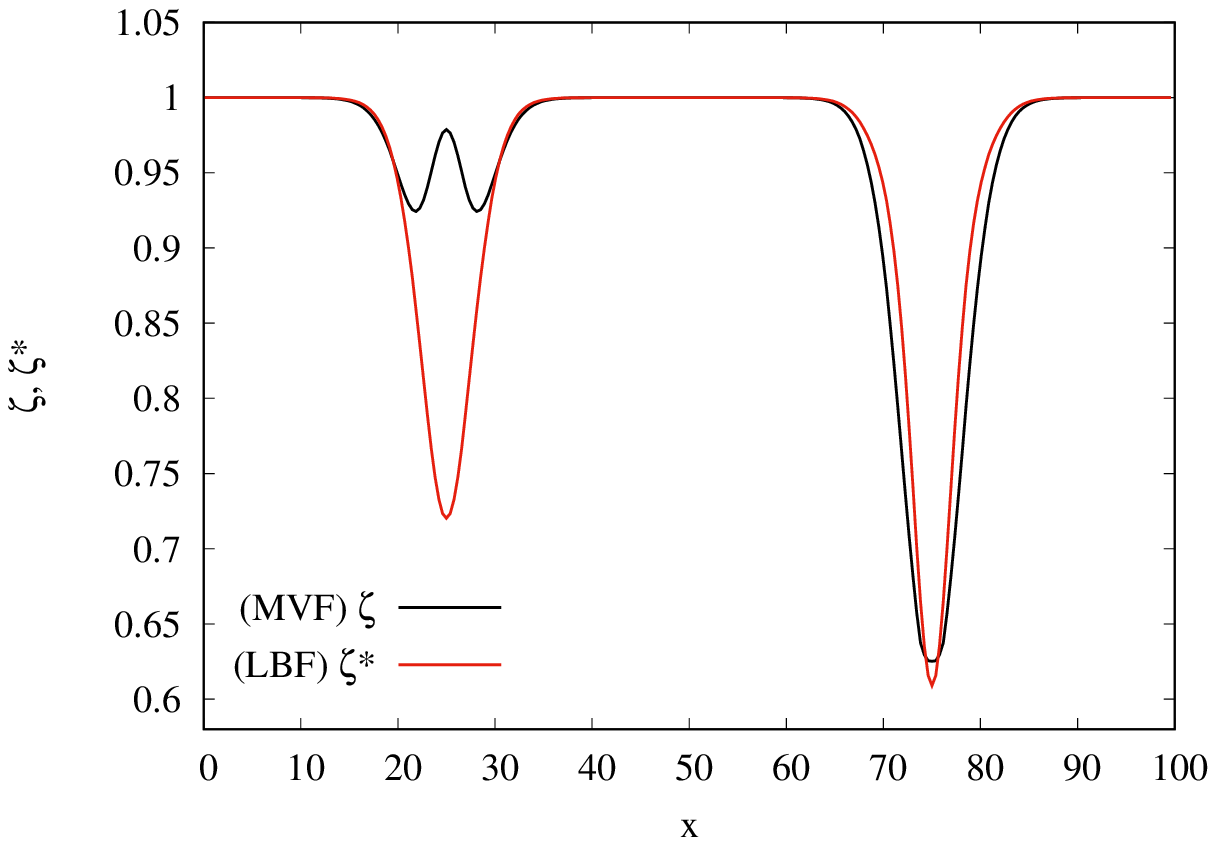}
\end{minipage}
\caption{Upper panel: profiles of anisotropy parameters $\eta$ (black line) and $\eta^*$ (red line); lower panel: profiles of agyrotropy parameters $\zeta$ (black line) and $\zeta^*$ (red line). Both panels refer to $\Delta \Phi=\pi/4$ in the nonuniform ${\bf B}$ case.
\label{fig:anisagirC2}}
\end{figure} 

\begin{figure}
\resizebox{\hsize}{!}{\includegraphics{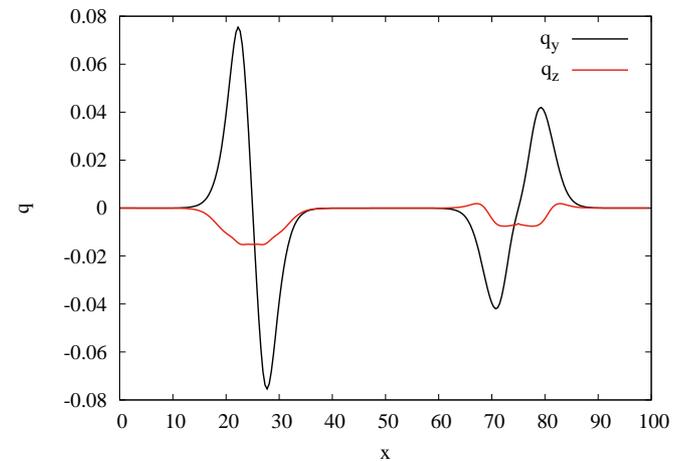}}
\caption{Profiles of heat flux components $q_y$ (black line) and $q_z$ (red line) are plotted as functions of $x$, calculated for $\Delta \Phi=\pi/4$ in the case of nonuniform ${\bf B}$.
\label{fig:calC2}}
\end{figure} 

\begin{figure*}[ht]
\begin{minipage}[ht]{0.32\textwidth}
\centering
  \includegraphics[width=\textwidth]{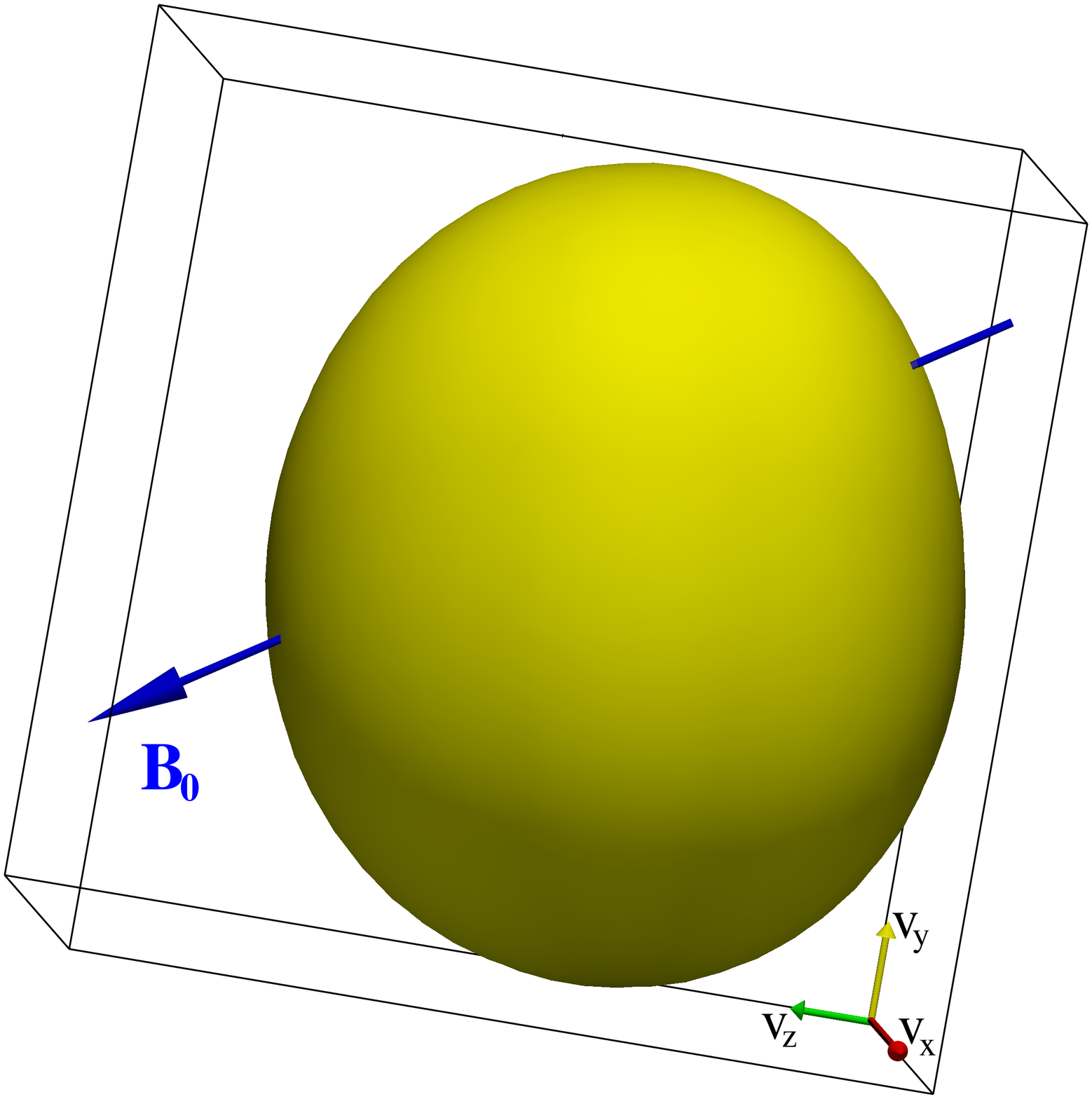}
\end{minipage}
\begin{minipage}[ht]{0.32\textwidth}
   \centering
   \includegraphics[width=\textwidth]{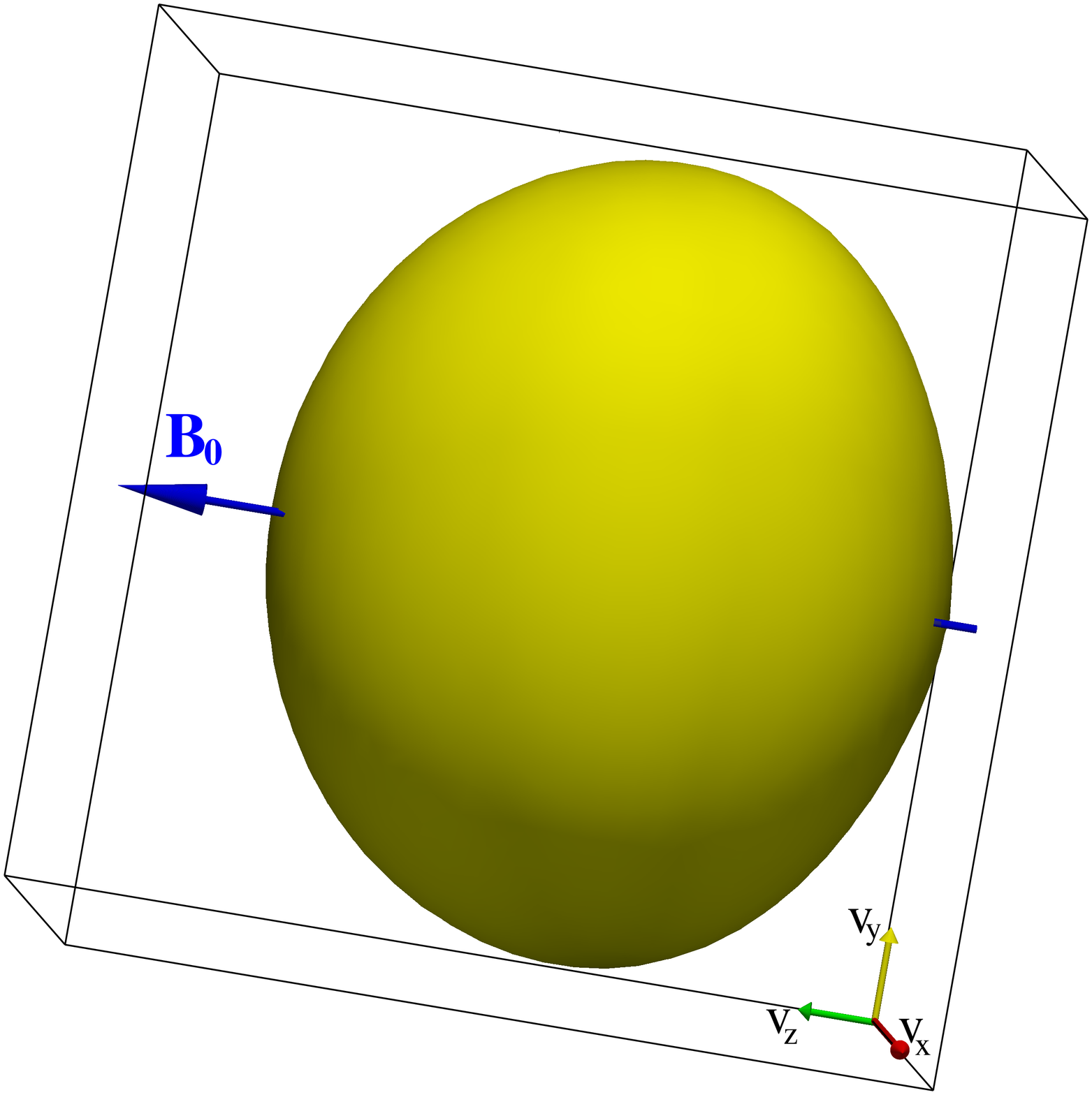}
\end{minipage} 
\begin{minipage}[ht]{0.32\textwidth}
   \centering
   \includegraphics[width=\textwidth]{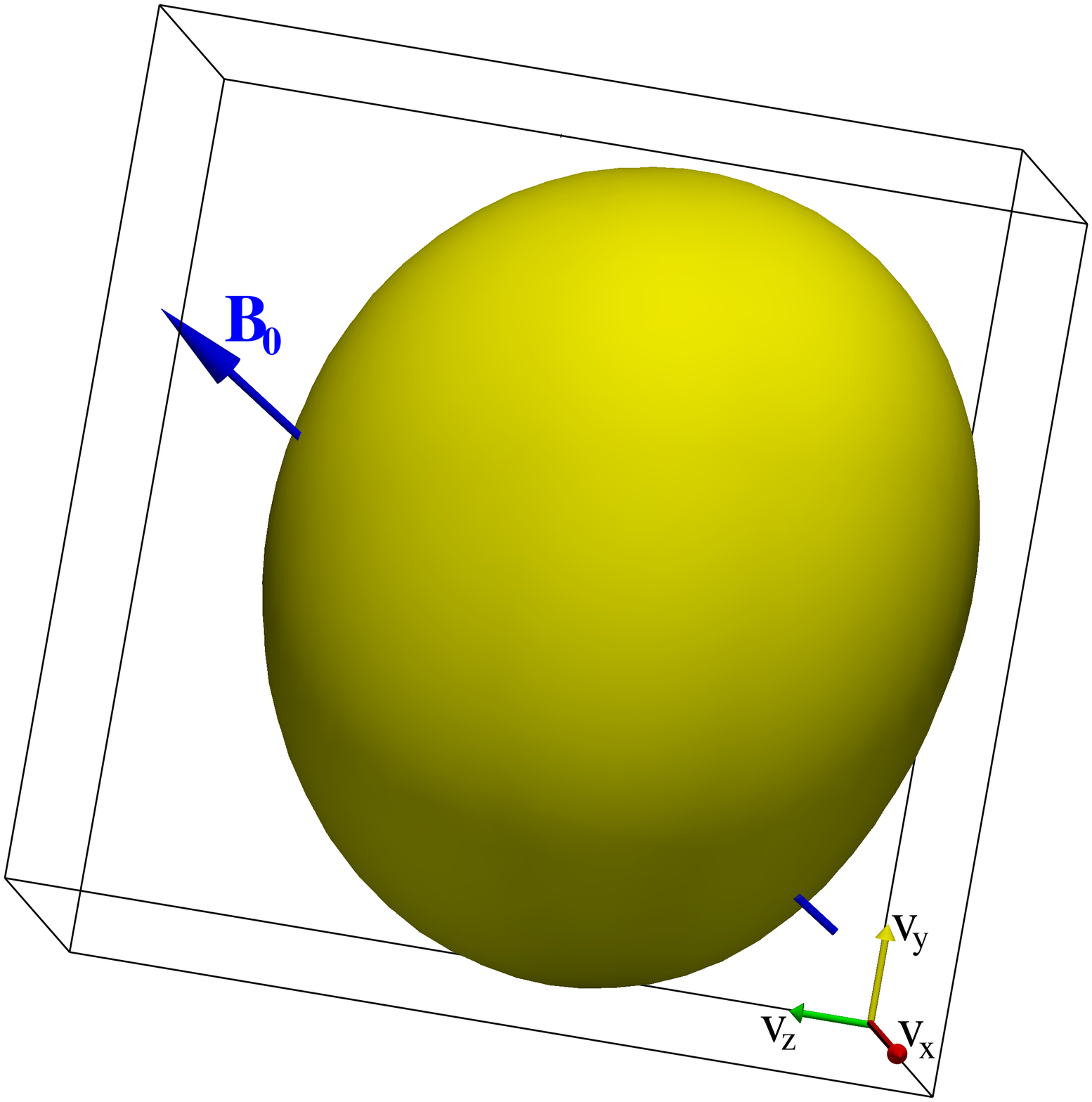} 
 \end{minipage}
\caption{Iso-surface plots of proton DFs in three different positions along the $x$-axis: $x=L/4-2.34$ (left panel),$x=L/4$ (middle panel), $x=L/4+2.34$ (right panel). The blue arrow indicates the direction of the background magnetic field ($B_0$).}
\label{fig:fd3D}
\end{figure*}

In Fig. \ref{fig:velC2} the profiles of the bulk velocity components $u_y(x)$ and $u_z(x)$ are plotted ($u_x$ is vanishing). The bulk velocity is not exactly aligned along $y$; however, it is $|u_z(x)| \ll |u_y(x)|$. Therefore, the desired condition that ${\bf u}$ is along the same direction in the whole domain is reasonably well satisfied by the DF (\ref{fC2}). In Fig.s \ref{fig:anisagirC2} the anisotropy $\eta$, $\eta^*$ and agyrotropy $\zeta$, $\zeta^*$ parameters are plotted in the case $\Delta\Phi=\pi/4$. The DF is anisotropic and agyrotropic at the shear layers, with respect to both MVF and LBF. Profiles of heat flux components $q_y$ and $q_z$ are plotted in Fig. \ref{fig:calC2} in the case $\Delta \Phi=\pi/4$ (it is $q_x=0$). A peculiar feature is that the $q_z$ component averaged across one shear layer $\langle q_z\rangle_{\rm sl}$ is no longer vanishing; this holds for both shear layers where a net heat flux directed opposite to the mean magnetic field $\langle {\bf B} \rangle_{\rm sl}$ has been found. A similar behaviour has been found also for other values of $\Delta \Phi$ (not shown). Finally, in Fig.\ref{fig:fd3D} 3D contour plots of the DF in the velocity space are plotted at three different positions across one shear layer. Blue arrows indicate the local direction of ${\bf B}$. Lack of both isotropy and gyrotropy is clearly visible. Moreover, in the position $x=L/4$ corresponding to the shear layer center (middle panel), an asymmetry with respect to the exchange $v_x\rightleftharpoons (-v_x)$ is evident. Such a feature is responsible for the net heat flux in the $-z$ direction.


\section{Conclusions}\label{section6}

In this paper we have derived exact solutions for the system of HVM equations representing a stationary shearing flow in collisionless magnetized plasmas. In many natural contexts there are configurations where a plasma is characterized by shearing flows. A kinetic description is necessary whenever the shear width is of the order of kinetics scales, like, for instance, in the case of Earth’s magnetophause \citep{sckopke81,kivelson95,faganello17}. 

The interest of building up stationary solutions can be related to the problem of describing the propagation and evolution of waves in a plasma with a stable shearing flow. The interaction between waves and the background inhomogeneity associated with the shearing flow moves the wave energy toward small scales, where kinetic effects are more effective. Moreover, the presence of a shearing flow can generate wave coupling, with an energy transfer among different wave modes. It is clear that, in order to properly study wave propagation, it is necessary that the background structure remains stationary; otherwise, a time evolution intrinsic of the background state would superpose to waves, making difficult to single out the wave contribution in the overall time evolution. Another possible application of exact shearing flow solutions can be found in the study of KHI, which takes place in unstable shearing velocity configurations. In fact, only a stationary unperturbed configuration allows us to properly describe both linear and nonlinear stages of the instability, as shown by \citet{settino20}. Therefore, in both cases using an exact stationary distribution function is crucial.

Stationary solutions in various configurations have been derived in previous studies of the fully kinetic case, i.e., involving the full set of ion and electron Vlasov-Maxwell equations \citep{ganguli88,nishikawa88,cai90,mahajan00}. However, the fully kinetic treatment is quite complex and such solutions have rarely been employed in numerical simulations like, for instance, in investigations of KHI. In this respect, the set of HVM equations represents a good compromise, because it correctly describes a plasma at scales of the order of or larger than ion scales but avoiding the complexity of a fully kinetic treatment. In this framework, in Paper I exactly stationary solutions have been found in the special cases where the magnetic field is uniform and either parallel or perpendicular to the bulk velocity. In the present paper we have extended these previous results to more general magnetized shearing flow configurations: 1) uniform obliquely-directed magnetic field and 2) variable-directed uniform-intensity magnetic field. In the first configuration the magnetic field ${\bf B}$ is uniform and forms a constant angle  with the bulk velocity ${\bf u}$. In the second configuration the angle between ${\bf B}$ and ${\bf u}$ is not constant and changes along the direction perpendicular to ${\bf u}$ and ${\bf B}$. Such configurations are more general than those considered in Paper I and they can more realistically represent the magnetic configuration across the magnetopause \citep[e.g.,][]{faganello17}. Therefore, they can be employed as a background state in future studies of KHI in the magnetopause in various magnetic configurations, as we are planning to do.  

Stationary DFs have been built up as appropriate combinations of single-particle constants of motions, that have been derived through successive generalizations of those found in Paper I. Stationarity with respect to the whole set of HVM equations has been verified. Explicit analytical forms for the DFs has been derived within the local approximation where both $E(x)$ and $B(x)$ have linear profiles. For more general profiles a numerical procedure has been set up which integrate single-particle motion, calculate the corresponding values of involved constants of motions and derive the DF on a grid in the phase space. We characterized our results calculating profiles of moments such as density, bulk velocity, heat flux, as well as quantities quantifying departures of the DF from Maxwellianity. In particular, we found that in the shear layers the DF is characterized by a marked anisotropy and agyrotropy. A peculiar property found in the shear regions is a net heat flux in the direction antiparallel to the mean ${\bf B}$ due to an asymmetry of the DF. We also verified that the obtained bulk velocity presents the desired orientation and planarity to a very good extent in both cases. A numerical simulation performed by means of the HVM code has allowed us to verify that departures from stationarity due to numerical errors remain to a low level, at least in the uniform ${\bf B}$ case.

Further possible developments of the present work that we are planning to do involve the inclusion of inhomogeneities of magnetic field intensity $|{\bf B}|$. This would allow us to perform a further step toward a more realistic representation of configurations like those found around the magnetopause. We also plan to use the configurations here considered to simulate the development of KHI in situations where kinetic effects at ion scales can play an important role.

\begin{acknowledgements}
     Numerical simulations have been run on Marconi supercomputer at CINECA (Italy) within the ISCRA projects: IsC68\_TURB-KHI and IsB19\_6DVLAIDA. This work has received funding from the European Unions Horizon 2020 research and innovation programme under grant agreement no. 776262 (AIDA, www.aida-space.eu)
\end{acknowledgements}

%
%

%

%

\begin{appendix} 
\section{Local approximation for uniform oblique ${\bf B}$.}\label{App:A}
In this appendix we study in more detail a configuration where the electric field has the linear profile (\ref{Elin}), that can be considered as a local approximation for a general profile $E(x)$ around the position $x_0$. In particular, we derive the expressions for the guiding center velocity and for the DF [Eq. (\ref{fC1la})]. The form for the magnetic field is given by the Eq.s (\ref{EB}) with $\Phi(x)=\Phi_0={\rm const}$. 
With respect to the reference frame $S'$, the normalized vector potential has the form ${\bf a}(x)=x {\bf e}_{y'}$
In this case, Eq. (\ref{1Dmot}) becomes $d^2 x/dt^2+\omega^2 x=(e/m) \left( E_0-\alpha_0 x_0\right) + \Omega_p w_{y'}$
where $\omega^2 = \Omega_p^2 - e\alpha_0/m$ and $w_{y'}=v_{y'}+\Omega_p x$ is a constant of motion. If $\alpha_0 < m\Omega_p^2/m$, the above equation describes an harmonic oscillator of solution 
\begin{equation}\label{xA}
x(t) = R_0 \sin \left( \omega t + \varphi \right) + x_c
\end{equation}
where $R_0$ and $\varphi$ are the amplitude and the phase of the oscillation. The guiding center position is given by the constant term:
\begin{equation}\label{xcA}
x_c=\langle x \rangle_\tau = \frac{1}{\omega^2} \left[ \frac{e}{m} \left( E_0-\alpha_0 x_0\right) + \Omega_p w_{y'} \right]
\end{equation}
that can be expressed as a function of $x$ and $v_{y'}$:
\begin{equation}\label{xc2A}
x_c=\frac{1}{1-c\alpha_0/(\Omega_p B_0)} \left[ x - \frac{v_{y'}}{\Omega_p}+\frac{c}{\Omega_p B_0} \left(E_0 - \alpha_0 x_0 \right)\right]
\end{equation}

The velocity components are given by
\begin{equation}\label{vxA}
v_x(t)=R_0 \omega \cos \left( \omega t + \varphi\right)
\end{equation}
\begin{equation}\label{vyA}
v_{y'}(t)=-\Omega_p x(t)+w_{y'} = -R_0\Omega_p \sin \left( \omega t + \varphi \right) + v_{c\perp}
\end{equation}
and $v_{z'}(t)={\rm const}$, where $v_{c\perp}=\langle v_{y'}(t)\rangle_\tau = -\Omega_p x_c + w_{y'}$. Using Eq. (\ref{xcA}) into this expression we obtain $v_{c\perp} = -\left[ e/(m\omega^2)\right] \left( \Omega_p E_0 + \alpha_0 v_{0y'} \right)$
where $v_{0y'}=-\Omega_p x_0 + w_{y'}$ is the value of $v_{y'}$ at the position $x_0$. On the other hand, being $w_{y'}$ a constant of motion, it is 
$v_{0y'}=v_{c\perp}+\Omega_p (x_c-x_0)$ [see Eq. (\ref{wywz})]. Using this relation, from the above equations we derive:
\begin{equation}\label{vc2A}
v_{c\perp}=-\frac{e}{m\omega} \left[ E_0 + \alpha_0(x_c-x_0)\right] = -c\frac{E(x_c)}{B_0}
\end{equation}
indicating that the guiding center velocity component perpendicular to ${\bf B}$ is given by the ${\bf E}\times{\bf B}$ drift velocity calculated at the guiding center position $x_c$. 

The argument of the exponential in Eq. (\ref{fC1}) contains the constant of motion $2\mathcal{E}_\perp/m-v_{c\perp}^2=v_{x}^2+v_{y'}^2+e\phi-v_{c\perp}^2$. Such a constant can be evaluated at $x=x_c$, where the potential is $\phi=0$; at that position it is $\sin (\omega t+\varphi)=0$ and then $v_x(x=x_c)=\pm R_0\omega$ and $v_{y'}(x=x_c)=v_{c\perp}$ [see Eq.s (\ref{vxA}), (\ref{vyA})]. As a consequence, it is $2\mathcal{E}_\perp/m-v_{c\perp}^2=R_0^2 \omega^2$ and Eq. (\ref{fC1}) can be written
\begin{equation}\label{fA}
f_{\Phi_0}^{la} = C \exp \left[ -\frac{R_0^2 \omega^2 + \left( v_{z'} - v_{c\perp} \tan\Phi_0\right)^2}{2v_{th,p}^2}\right]
\end{equation}
Using Eq.s (\ref{vxA}), (\ref{vyA}) and (\ref{vc2A}), the constant $R_0^2 \omega^2$ can be expressed as a function of $x$ , $v_{x}$ and $v_{y'}$ in the following form:
\begin{equation}\label{R02omega2}
R_0^2 \omega^2 = v_x^2 + \frac{\omega^2}{\left( \Omega_p - c\alpha_0/B_0\right)^2} \left( v_{y'} + \frac{cE(x)}{B_0}\right)^2
\end{equation}
From Eq.s (\ref{vc2A}) and (\ref{xc2A}) we express $v_{c\perp}$ as a function of $x$ and $v_{y'}$: 
\begin{equation}\label{vcperpA}
v_{c\perp}=v_{d0}-\frac{c\alpha_0}{\Omega_p B_0 - c\alpha_0}\left[ \Omega_p \left( x-x_0\right)+v_y-v_{d0}\right]
\end{equation}
where $v_{d0}=-cE_0/B_0$ is the drift velocity at the position $x_0$. Inserting equations (\ref{R02omega2}) and (\ref{vcperpA}) into Eq. (\ref{fA}), this can be manipulated and finally written in the form (\ref{fC1la}), where the normalization constant $C$ has been determined by imposing that $\int f_{\Phi_0}^{la} d^3 {\bf v}=n_0$. A straightforward calculation allows one to find Eq. (\ref{uC1loc}).

\section{Local approximation for uniform-intensity variable-direction ${\bf B}$.}\label{App:B}
In this appendix we derive the form for the DF in the case of the local approximation for a magnetic field with uniform intensity and variable direction. The electric field has the form (\ref{Elin}) and the magnetic field components have linear profiles given by Eq.s (\ref{Bylin}) and (\ref{Bzlin}), where $x_0$ represents the center of the Taylor expansion. The corresponding potentials are 
\begin{equation}\label{philin}
\phi(x)=-E_0 (x-x_0)-\frac{\alpha_0}{2}(x-x_0)^2 + \phi_0
\end{equation}
\begin{equation}\label{ayazlin}
a_y(x)=b_{0z}(x-x_0) - \frac{\gamma_z}{2}(x-x_0)^2 \;\; , \;\; a_z(x)=-b_{0y}(x-x_0) - \frac{\gamma_y}{2}(x-x_0)^2
\end{equation}
with $\phi_0={\rm const}$. Imposing the $|{\bf B}(x)|=B_0={\rm uniform}$, at the linear order we obtain the relations
\begin{equation}\label{b0gamma}
b_{0y}^2+b_{0z}^2=1 \;\; , \;\; b_{0y}\gamma_y = b_{0z}\gamma_z
\end{equation}	
Coherently with the local approximation, in what follows we will neglect all terms cubic or of higher order in the ratios $|\alpha_0(x-x_0)/E_0|$ and $|\gamma_{y,z}(x-x_0)/b_{0y,0z}|$. The single particle equation of motion in the $x$ direction has the form $d^2 x/dt^2 + \omega^2 x = (e/m)(E_0 - \alpha_0 x_0)+\Omega_p W_0$, where $\omega$ is given by Eq. (\ref{omega}) and 
\begin{equation}\label{W0}
W_0 = (w_y \gamma_z + w_z \gamma_y)x_0 + w_y b_{0z} - w_z b_{0y} + \Omega_p x_0
\end{equation}
If $\alpha_0 < (m/e)\left[ \Omega_p^2 + \Omega_p \left( w_y\gamma_z + w_z\gamma_y\right) \right]$ it is $\omega^2>0$ and the particle motion along $x$ is that of an harmonic oscillator 
\begin{equation}\label{xvxB}
x(t)=R_0 \sin(\omega t + \varphi) + x_c \;\; , \;\; v_x(t)=R_0 \omega \cos(\omega t + \varphi)
\end{equation}
where $R_0$ is the amplitude, $\varphi$ is the phase and $x_c$ is the center of motion:
\begin{eqnarray}\label{xcC2}
x_c =&&\frac{1}{\omega^2} \left\{ \frac{e}{m} \left( E_0+\alpha_0x_0\right) + \right. \nonumber \\
&&\left.+\Omega_p \left[ w_y \left( b_{0z} + \gamma_z x_0 \right) - w_z \left( b_{0y}-\gamma_y x_0 \right)+ \Omega x_0 \right] \right\}
\end{eqnarray}
corresponding to the $x$ position of the guiding center. Otherwise, if $\alpha_0 \ge (m/e)\left[ \Omega_p^2 + \Omega_p \left( w_y\gamma_z + w_z\gamma_y\right) \right]$ the motion along $x$ is not bounded and the local approximation breaks down. 
On the other hand, the motion equation along $x$ can be re-written in the form $dv_x(t)/dt=(e/m)E(x(t))+\Omega_p\left\{ v_y\left[da_y(x(t))/dx\right]+v_z\left[da_z(x(t))/dx\right]\right\}$. Inserting Eq.s (\ref{xvxB}) on both sides, and neglecting third-order terms we obtain the following relation:
\begin{equation}\label{compat}
-c E(x_c)=w_y B_z(x_c) - w_z B_y(x_c) - \Omega_p B_0 (x_c - x_0)
\end{equation}
Using Eq.s (\ref{wywz}), (\ref{ayazlin}) and (\ref{xvxB}) we derive expressions for $v_y$ and $v_z$:
\begin{equation}\label{vyB}
v_y(t) = w_y -\Omega_p a_y(x_c) - \frac{\Omega_p B_z(x_c)}{B_0} R_0 \sin(\omega t + \varphi) + \frac{\Omega_p \gamma_z}{2} R_0^2 \sin^2(\omega t + \varphi)
\end{equation}
\begin{equation}\label{vzB}
v_z(t) = w_z -\Omega_p a_z(x_c) + \frac{\Omega_p B_y(x_c)}{B_0} R_0 \sin(\omega t + \varphi) + \frac{\Omega_p \gamma_y}{2} R_0^2 \sin^2(\omega t + \varphi)
\end{equation}
We observe that, in consequence of the dependence on $\sin^2(\omega t+\varphi)$ in $v_y(t)$ and $v_z(t)$, particles trajectories in the velocity space are no longer ellipses but have a more complex form. The $y$ and $z$ components of the guiding center velocity are
\begin{equation}\label{vcyB}
v_{cy} = \langle v_y \rangle_\tau = \frac{\Omega_p \gamma_z}{2} (x_c-x_0)^2 - \Omega_p b_{0z}(x_c-x_0) + \frac{\Omega_p R_0^2 \gamma_z}{4} + w_y
\end{equation}
\begin{equation}\label{vczB}
v_{cz} = \langle v_z \rangle_\tau = \frac{\Omega_p \gamma_y}{2} (x_c-x_0)^2 + \Omega_p b_{0y}(x_c-x_0) + \frac{\Omega_p R_0^2 \gamma_y}{4} + w_z
\end{equation}
From these equations we derive expressions for $w_y$ and $w_z$ that are inserted into Eq. (\ref{compat}). Neglecting third-order terms, we obtain $v_{cz}B_y(x_c) - v_{cy}B_z(x_c) = cE(x_c)$, that determines the component of the guiding center velocity perpendicular to ${\bf B}(x_c)$:
\begin{equation}\label{vcperpB}
{\bf v}_{c\perp}=c\frac{{\bf E}(x_c) \times{\bf B}(x_c)}{B^2}
\end{equation}
which corresponds to the ${\bf E}\times {\bf B}$ drift velocity calculated at the guiding center position $x_c$; this is Eq. (\ref{vcperp}).

\end{appendix}
\end{document}